%% file: FL_IoT_Survey_Manuscript.tex
\documentclass[journal,10pt,twocolumn,twoside]{IEEEtran}

\usepackage{mathtools}
\usepackage{empheq}
\usepackage{algpseudocode, algorithm}
\usepackage{algorithmicx}
\usepackage{subfig}
\usepackage[table]{xcolor}%
\usepackage{varwidth}
\usepackage{url}
\usepackage{multirow}
\usepackage{amssymb}
\usepackage{mathtools}
\usepackage{changes}
\usepackage[short]{optidef}
\usepackage{caption}
\usepackage{amsmath}
\usepackage{textcomp}
\usepackage{gensymb}
\usepackage{cite}
\usepackage{booktabs}
\usepackage{hyperref}
\usepackage{balance}
\usepackage{booktabs}
\usepackage{array}
\usepackage{balance}
\newcolumntype{P}[1]{>{\centering\arraybackslash}p{#1}}
\usepackage{makecell}

\begin{document}
	
	\title{Federated Learning for Internet of Things:\\ A Comprehensive Survey}
	
	\author{Dinh C. Nguyen,	Ming Ding, Pubudu N. Pathirana,
		Aruna Seneviratne, \\Jun Li, and H. Vincent Poor,~\IEEEmembership{Fellow,~IEEE}
		
		\thanks{Dinh C. Nguyen is with the School of Engineering, Deakin University, Waurn Ponds, VIC 3216, Australia, and also with  Data61, CSIRO, Docklands, Melbourne, Australia  (e-mail: cdnguyen@deakin.edu.au).}
		\thanks{Ming Ding is with Data61, CSIRO, Australia (email: ming.ding@data61.csiro.au).}
		\thanks{Pubudu N. Pathirana is with the School of Engineering, Deakin University, Waurn Ponds, VIC 3216, Australia (email: pubudu.pathirana@deakin.edu.au).}		
		\thanks{Aruna Seneviratne is with the School of Electrical Engineering and Telecommunications, University of New South Wales (UNSW), NSW, Australia (e-mail: a.seneviratne@unsw.edu.au).}
		\thanks{Jun Li is with the School of Electrical and Optical Engineering, Nanjing University of Science and Technology, Nanjing 210094, China (e-mail: jun.li @njust.edu.cn).}
		\thanks{H. Vincent Poor is with the Department of Electrical and Computer Engineering, Princeton University, Princeton, NJ 08544 USA (e-mail: poor@princeton.edu).}
		\thanks{This work was supported in part by the CSIRO Data61, Australia, and in part by U.S. National Science Foundation under Grant CCF-1908308. The work of Jun Li was supported by National Natural Science Foundation of China under Grant 61872184. }
	}
	
	\markboth{Accepted at IEEE Communications Surveys \& Tutorials}%
	{}
	
\maketitle
\IEEEdisplaynontitleabstractindextext
\IEEEpeerreviewmaketitle

	\begin{abstract}
	The Internet of Things (IoT) is penetrating many facets of our daily life with the proliferation of intelligent services and applications empowered by artificial intelligence (AI). Traditionally, AI techniques require centralized data collection and processing that may not be feasible in realistic application scenarios due to the high scalability of modern IoT networks and growing data privacy concerns. Federated Learning (FL) has emerged as a distributed collaborative AI approach that can enable many intelligent IoT applications, by allowing for AI training at distributed IoT devices without the need for data sharing.  In this article, we provide a comprehensive  survey of the emerging applications of FL in IoT networks, beginning from an introduction to the recent advances in FL and IoT to a discussion of their integration. Particularly, we explore and analyze the potential of FL for enabling a wide range of IoT services, including IoT data sharing, data offloading and caching, attack detection, localization, mobile crowdsensing, and IoT privacy and security. We then provide an extensive survey of the use of FL in various key IoT applications such as smart healthcare, smart transportation, Unmanned Aerial Vehicles (UAVs), smart cities, and smart industry. The important lessons learned from this review of the FL-IoT services and applications are also highlighted. We complete this survey by highlighting the current challenges and possible directions for future research in this booming area. 
	\end{abstract}
	

	\input{Introduction.tex}

	\input{State-of-Art.tex}

	\input{FL_Services.tex}

	\input{FL_Applications.tex}
	\input{Lessons_Learned.tex}
	\input{Challenges_Future-Directions.tex}
	
	\input{Conclusion.tex}
	\balance
	\bibliography{Ref}
	\bibliographystyle{IEEEtran}

\end{document}

%% file: Introduction.tex
\section{Introduction}
\label{Sec:Introduction}
Recent years have witnessed the rapid development of the Internet of Things (IoT) which provides ubiquitous sensing and computing capabilities to connect a broad range of things to the Internet \cite{1}. To obtain insights into data generated from ubiquitous IoT devices, artificial intelligence (AI) techniques {such as deep learning (DL)} have been widely exploited to train data models for enabling intelligent IoT applications such as smart healthcare, smart transportation, and smart city \cite{2}. Traditionally, AI functions are placed in a cloud server or a data center for data learning and modeling \cite{3}, which incurs critical limitations given the IoT data explosion.  {According to Cisco, there will be nearly 850 ZB of data generated by all people, machines, and things at the network edge by 2021. In sharp contrast, the global data center traffic will only reach 20.6 ZB in this year \cite{Cisco}.} With such a tremendous growth of IoT data at the network edge, the offloading of massive IoT data to the remote servers may be infeasible do to the required network resources and the incurred latency. The use of third-party servers for AI training also raises privacy concerns such as data breaches as the training data may contain sensitive information such as user addresses or personal preferences \cite{4}. It is thus highly necessary for developing innovative AI approaches to realize efficient and privacy-enhanced intelligent IoT networks and applications. 

Recently, the concept of federated learning (FL) has been proposed for building intelligent and privacy-enhanced IoT systems. Technically, FL is a distributed collaborative AI approach that allows for data training by coordinating multiple devices with a central server without sharing actual datasets \cite{5}. For instance, multiple IoT devices can act as workers to communicate with an aggregator (e.g., a server) for performing neural network training in intelligent IoT networks.  More specifically, the aggregator first initiates a global model with learning parameters. Each worker downloads the current model from the aggregator, computes its model update, e.g., via stochastic gradient descent (SGD), by using its local dataset, and offloads the computed local update to the aggregator. Then, the aggregator combines all local model updates and constructs a new improved global model. By using the computing power of distributed workers, the aggregator can enhance the training quality {while minimizing user privacy leakage}. Finally, the local workers download the global update from the aggregator, and compute their next local update until the global training is complete. With its innovative  operational concept, FL can offer various important benefits for IoT applications as follows:
\begin{itemize}
	\item 	\textit{Data Privacy Enhancement:} In FL, the raw data are not required for the training at the aggregator. {Therefore, the leakage of sensitive user information to the external third-party is minimized and a degree of data privacy is provided.}  Following the increasingly stringent data privacy protection legislation such as the General Data Protection Regulation (GDPR) \cite{7}, the privacy protection feature makes FL an ideal solution for building intelligent and safe IoT systems.  
	\item	\textit{Low-latency Network Communication:} Since there is no requirement for transmitting IoT data to the server, the use of FL helps reduce communication latencies caused by data offloading. In return, it also saves network resources, e.g., spectrum and transmit power, in the data training. 
	\item	\textit{Enhanced Learning Quality:} By attracting much computation resources and diverse datasets from a network of IoT devices, FL has the potential to enhance the convergence rate of the overall training process and achieve better learning accuracy rates \cite{FLchain}, which might not be achieved by using centralized AI approaches with insufficient data and constrained computational capabilities. In return, FL also improves the scalability of intelligent networks due to its distributed learning nature.
\end{itemize}
With these unique advantages, FL has been proposed for use in a variety of IoT applications, such as smart healthcare, smart transportation, Unmanned Aerial Vehicles (UAVs), etc. For example, FL has facilitated smart health services by enabling machine learning (ML) modeling without sharing patient data across multiple medical institutions \cite{8}. By using FL, health data owners, e.g., hospitals, do not need to exchange their healthcare records with each other; instead, they train the AI model locally and only upload the trained parameters to the aggregator for global computation. In this way, FL creates collaborative healthcare environments among different hospitals to accelerate patient diagnosis and treatment, without sacrificing user privacy. In transportation systems, FL has also proved its potential to provide smart vehicular services \cite{112}, such as autonomous driving, road safety prediction, vehicular detection with high training accuracy and privacy enhancement, by coordinating vehicles with roadside units for collaborative data learning.  The success of recent FL-IoT applications makes now the right time to {draw attention to this prominent area of research.}
\subsection{Comparison and Our Contributions}
Driven by the recent advances of FL and IoT,  several reviews of related work have appeared. For example, the study in \cite{10} provided a survey of the FL concept and analyzed its key system components such as data distribution, machine learning model, privacy mechanism, and communication architecture. Similarly, the authors in \cite{11} discussed the FL concept from the architecture perspective, along with the analysis of basic applications of FL in business. Other articles in \cite{cp0, cp1, cp5} also presented the concept of FL architectures, software, platforms, and protocols and discussed a few possible research challenges in FL deployments. Meanwhile, the study in \cite{12} analyzed technical challenges in FL system, including data privacy bottlenecks and network attacks. Moreover, the use of FL in mobile edge networks was investigated in \cite{13}. The key focuses are on the discussion of the challenges in FL implementation in edge networks and the roles of FL in edge network optimization. Recently, the potential of FL in wireless networks has been also explored from various aspects. For example, an overview on the use of FL in fog radio access networks was presented in \cite{14}, with a discussion of the fundamental FL theory with respect to the accuracy loss correction and the model compression. An overview on the use of FL for wireless communications was provided in \cite{15} where the roles of FL in 5G applications are discussed, such as edge computing, spectrum sharing, and 5G core network management. The integration of FL and 6G wireless applications was explored in \cite{16}, where the key challenges of using FL for 6G networks are highlighted. Furthermore, a survey in \cite{17} paid attention to the discussion of the data preservation methods applied to FL to protect users in the FL training. Another work in \cite{cp2} presented a survey on the key architectures of FL models with a very brief introduction to the use of FL in health informatics. The potential of FL in vehicular networks was explored in \cite{cp3}, and the integrated FL-UAVs models with representative use cases were discussed in \cite{cp4}. The comparison of the related works and our paper is summarized in Table~\ref{Table:Comparisons}.
\begin{table*}
	\centering
	\caption{\textcolor{black}{Existing surveys on FL-related topics and our new contributions. }}
	\label{Table:Comparisons}
	{\color{black}
		\setlength{\tabcolsep}{5pt}
		\begin{tabular}{|P{1.3cm}|P{2cm}|P{7.8cm}|P{5.5cm}|}
			\hline
			\centering \textbf{Related works}& 
			\centering \textbf{Topic}&	
			\textbf{Key contributions}&	
			\textbf{Limitations}
			\\
			\hline
			\cite{10}&	FL concept&	A survey on the FL system components, such as data distribution, machine learning model, privacy mechanism, and communication architecture.&The applications of FL in IoT networks have not been presented.
			\\
			\hline
			\cite{11} &	FL concept &	A survey on the FL concept with a basic introduction to definitions, architectures and applications of FL.&The applications of FL in IoT networks have not been presented.
			\\
			\hline
			\cite{cp0}&	FL concept&	A survey on the architectures,  algorithms,  and  data  processing  methods in FL-based wireless networks.&The paper only analyzes the use of FL in wireless networks, while its roles in IoT have not been presented.
			\\
			\hline
			\cite{cp1}&	FL concept&	A survey on the FL hardware, software, platforms, and protocols with limited discussion of FL-based healthcare use cases. &The analysis of FL-healthcare is limited. Moreover,  discussions for other IoT domains are lacked. 
			\\
			\hline
			\cite{cp5}&	FL concept&	A systematic survey on the FL concept, research challenges and solutions from the perspective of software engineering. &The use of FL in IoT services and applications has not been presented.
			\\
			\hline
			\cite{12} &	FL and data privacy &	A brief overview on the data privacy challenges and attacks in FL systems, along with possible solutions for data privacy protection in FL designs.&The paper only focuses on privacy aspects in FL. 
			\\
			\hline
			\cite{13} &	FL and mobile edge networks &	A survey on the integration of FL and edge computing, including the discussion of the challenges in FL implementation in edge networks and the analysis the use of FL in edge network optimization.  &The applications of FL in IoT networks and services have not been explored and discussed. 
			\\
			\hline
			\cite{14} &	FL and fog radio access networks &	An overview on the use of FL in fog radio access networks, with a discussion of the fundamental FL theory with respect to the accuracy loss correction and the model compression. &The applications of FL in IoT networks have not been presented.
			\\
			\hline
			\cite{15} &	FL and wireless communication &	An overview on the use of FL for wireless communications and the roles of FL in 5G applications such as edge computing, spectrum sharing, and 5G core network management. &The benefits of FL in IoT services such as IoT data sharing data offloading, localization have not been explored. 
			\\
			\hline
			\cite{16} &	FL and 6G communication &	A survey on the integration of FL and 6G applications, along with a discussion of the challenges in the use of FL for 6G networks.&The discussion of the FL applications for IoT networks has not been provided.
			\\
			\hline
			\cite{17} &	FL and data privacy in IoT analytics &	A survey on the data preservation methods applied to FL to protect users in the FL training. &The paper only discusses on the privacy aspect of FL-IoT while a comprehensive survey on the FL-IoT integration is missing. 
			\\
			\hline
			\cite{cp2}&	FL and health informatics&	A survey on the concept and key architectures of FL models with a very brief introduction to the use of FL in health informatics. &The paper only focuses on discussing the roles of FL in the healthcare domain.
			\\
			\hline
			\cite{cp3}&	FL and vehicular networks&	A survey on the advances of FL and its application in vehicular networks.&The paper only focuses on discussing the roles of FL in the vehicular network and transportation domain.
			\\
			\hline
			\cite{cp4}&	FL and UAVs&	A brief survey on the application of FL in UAVs networks. &Only analysis on the FL-UAV is provided while other domains such as smart city, smart industry are not considered. 
			\\
			\hline
			\textit{Our paper} &	FL and IoT &	An extensive survey on the FL-IoT integration. Particularly,
			\begin{itemize}
				\item 	We extensively discuss the roles of FL in various key IoT services, i.e., IoT data sharing, data offloading and caching, attack detection, localization, mobile crowdsensing, and IoT privacy and security.
				\item 	The use of FL in important IoT applications is also analyzed in detail, including smart healthcare, smart transportation, UAVs, smart city, and smart industry.
				\item 	Taxonomy tables and key lessons learned are provided to provide insights into FL-IoT integration. Research challenges and directions are also highlighted. 
			\end{itemize} &-
			\\
			\hline
	\end{tabular}}
\end{table*}

\textcolor{black}{Although FL has been studied extensively in the literature, there is no existing work to provide a comprehensive and dedicated review of the use of FL in IoT networks and applications, {to the best of our knowledge}. The potential of FL in IoT services, such as IoT data sharing, data offloading, localization, etc., has not been explored in the open literature \cite{10,11,cp0,cp1,cp5}, \cite{13}. Moreover, a holistic discussion of the integrated FL-IoT applications from smart transportation to smart city is still missing \cite{cp2,cp3,cp4}. These limitations motivate us to conduct a more comprehensive review of the integration of FL in IoT networks. Particularly, we provide a state-of-the-art survey of the applications of FL in various key IoT services such as IoT data sharing, data offloading and caching, attack detection, localization, mobile crowdsensing, and IoT privacy and security.  The key contribution of this paper lies in the extensive discussion of the use of FL in a wide range of IoT applications, including smart healthcare, smart transportation, UAVs, smart city, and smart industry. The key lessons learned from the survey are also given. Finally, we discuss a number of important research challenges and highlight interesting future directions in FL-IoT.  To this end, the key contributions of this article are highlighted as follows:
\begin{enumerate}
\item We present a state-of-the-art survey on the application of FL in IoT networks, starting from an introduction to the recent advances in FL and IoT and the discussion of the visions behind their integration. 
\item	We discuss the opportunities created by FL in many key IoT services, namely IoT data sharing, data offloading and caching, attack detection, localization, mobile crowdsensing, and IoT privacy and security.
\item	We perform a holistic investigation and analysis of the potential of FL in a number of IoT applications, including smart healthcare, smart transportation, UAVs, smart city, and smart industry with discussions on communications and networking aspects. Taxonomy tables to summarize the key technical aspects, contributions and limitations of each FL approach used in IoT are also provided. 
\item	From the survey of FL-IoT services and applications, the key lessons learned are highlighted. Lastly, we identify a couple of important research challenges and then discuss possible directions for future research toward the full realization of FL-IoT.
\end{enumerate}}
\subsection{Structure of The Survey}

\begin{figure*}
	\centering
	\includegraphics[width=0.99\linewidth]{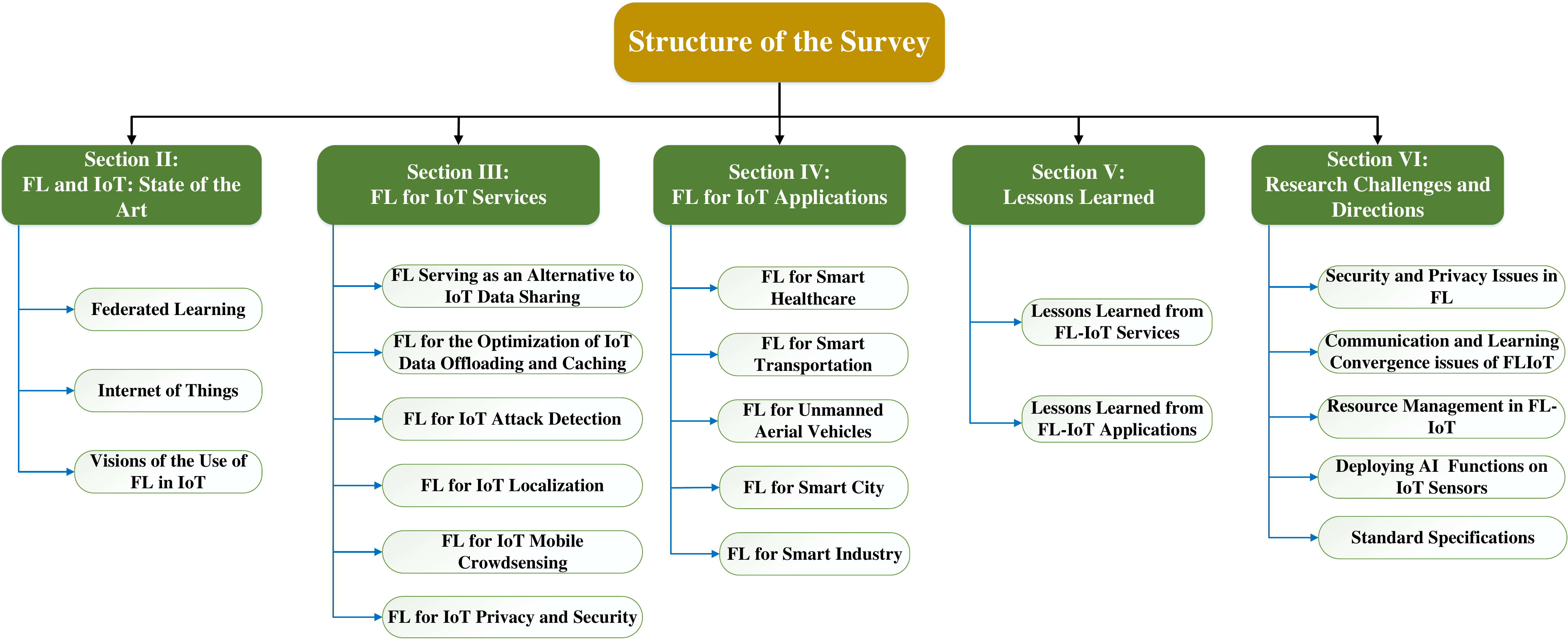}
	\caption{\textcolor{black}{Organization of this article.} }
	\label{Fig:FL_Application_Structure}
	\vspace{-0.1in}
\end{figure*}

This survey is organized as shown in Fig.~\ref{Fig:FL_Application_Structure}.  Section~\ref{Sec:State-of-Art} discusses the state-of-the-art of FL and IoT, and  the visions of their integration are also highlighted. We provide an extensive discussion of the use of FL in important IoT services in Section~\ref{Sec:FL_Services}, including IoT data sharing, data offloading and caching, attack detection, localization, and mobile crowdsensing. The opportunities brought by FL in a number of key IoT applications, such as smart healthcare, smart transportation, UAVs, smart city, and smart industry, are then explored and analyzed in Section~\ref{Sec:FL_Applications}. From the comprehensive survey, we summarize and highlight several key lessons learned in Section~\ref{Sec:Lessons_Learned}. Section~\ref {Sec:Challenges_Future-Directions} discusses the key research challenges, including threats in FL, performance issues of FL, resource management in FL systems, and heterogeneity issues in IoT networks, along with  possible directions for future research. Finally, Section~\ref{Sec:Conclusion} concludes the article. A list of key acronyms and abbreviations used throughout the paper is given in Table~\ref{Table:Acronyms}.
\begin{table}
	\caption{List of key acronyms.}
	\label{Table:Acronyms}
	\scriptsize
	\centering
	\captionsetup{font=scriptsize}
	\setlength{\tabcolsep}{5pt}
	\begin{tabular}{P{1.5cm}|p{3.6cm}}
		\hline
		\textbf{Acronyms}& 
		\textbf{Definitions}
		\\
		\hline
		FL& Federated Learning
		\\
		IoT& Internet of Things
		\\
		HFL &	Horizontal Federated Learning
		\\
		VFL&	Vertical Federated Learning 
		\\
		FTL&	Federated Transfer Learning
		\\
		CFL&	Centralized Federated Learning
		\\
		DFL&	Decentralized Federated Learning
		\\
		AI& Artificial Intelligence
		\\
		ML & Machine Learning
		\\
		DL & Deep Learning
		\\
		DRL & Deep Reinforcement Learning 
		\\
		SVM &Support Vector Machine
		\\
		NN &Neural Network
		\\
		DNN &Deep Neural Network
		\\
		CNN & Convolutional Neural Network
		\\
		RNN & Recurrent Neural Networks
		\\
		LSTM & Long-short Term Memory
		\\
		MEC & Mobile Edge Computing 
		\\
		UAV &  Unmanned Aerial Vehicle
		\\
		BS & Base Station
		\\
		V2V & Vehicle-to-Vehicle
		\\
		RSU & Roadside Unit 
		\\
		EHR & Electronic Health Record
		\\
		{Non-IID Data}& {Nonindependent and Identically Distributed Data}
		\\
		\hline
	\end{tabular}
	\label{tab1}
\end{table}

%% file: State-of-Art.tex
\section{FL and IoT: State of the Art}
\label{Sec:State-of-Art}
In this section, we present the state-of-the-art of FL and IoT. The visions of their integration are also discussed.  
\subsection{Federated Learning}
Since its inception in 2016 \cite{18}, FL has transformed many intelligent IoT applications by offering new AI solutions with its distributed and {privacy-enhancing} nature. The arrival of this emerging distributed AI technology has the potential to reshape the current intelligent IoT systems with advanced FL architectures. Following the recent advances in mobile hardware and the growing concerns of privacy leakage, FL is particularly attractive for building distributed IoT systems, by pushing AI functions, e.g., AI data training, to the network edge at IoT devices where the data reside. As a result, the user data are never shared directly with the third party while enabling the cooperative training of a shared global model, which benefits both network operators and IoT users in terms of network resource savings and privacy enhancement. FL thus would be a strong alternative for traditional centralized AI approaches and helps accelerate the deployment of  IoT services and applications at a larger scale. We here introduce the key concept of FL and then present some important FL categories used in IoT networks. 
\subsubsection{Key FL Concept}

The FL concept in IoT networks is composed of two main entities: the data clients, e.g., IoT devices and an aggregation server located at a base station (BS) or an access point (AP), as illustrated in Fig.~\ref{Fig:FL_Concept}. Let $\mathcal{K}  = \{1,2,..., K\}$ denote the set of participants who use IoT devices such as smartphone, laptop or tablet to collaboratively implement an FL algorithm for performing an IoT task. For example, in an IoT-based vehicular network \cite{115,116}, vehicles can join a shared FL process to sense the road traffic environment and produce a comprehensive traffic routing map for reducing traffic congestion. In the next generation of IoT networks, FL is of paramount importance for realizing full intelligence in IoT systems at the network edge, since a BS cannot collect all data from distributed IoT devices for AI/ML training. FL allows IoT users and the BS to train a shared global model while the raw data are remained at users’ devices. In an FL process, each IoT user $k$ participates in training a shared AI/ML model by using their own dataset $D_{k \in \mathcal{K}}$. Hereinafter, the FL model trained at the IoT device is called the \textit{local model} $ \textbf{w}_k$. After local training, IoT users upload their local model updates to the BS that then aggregates to build a shared model, called the \textit{global model} $w_G$. By relying on the distributed data training at the IoT devices, the aggregation server at the BS can enrich the training performance {without significantly compromising user data privacy}. As shown in Fig.~\ref{Fig:FL_Concept}, the general FL process includes the following key steps:
\begin{figure*}
	\centering
	\includegraphics[width=0.98\linewidth]{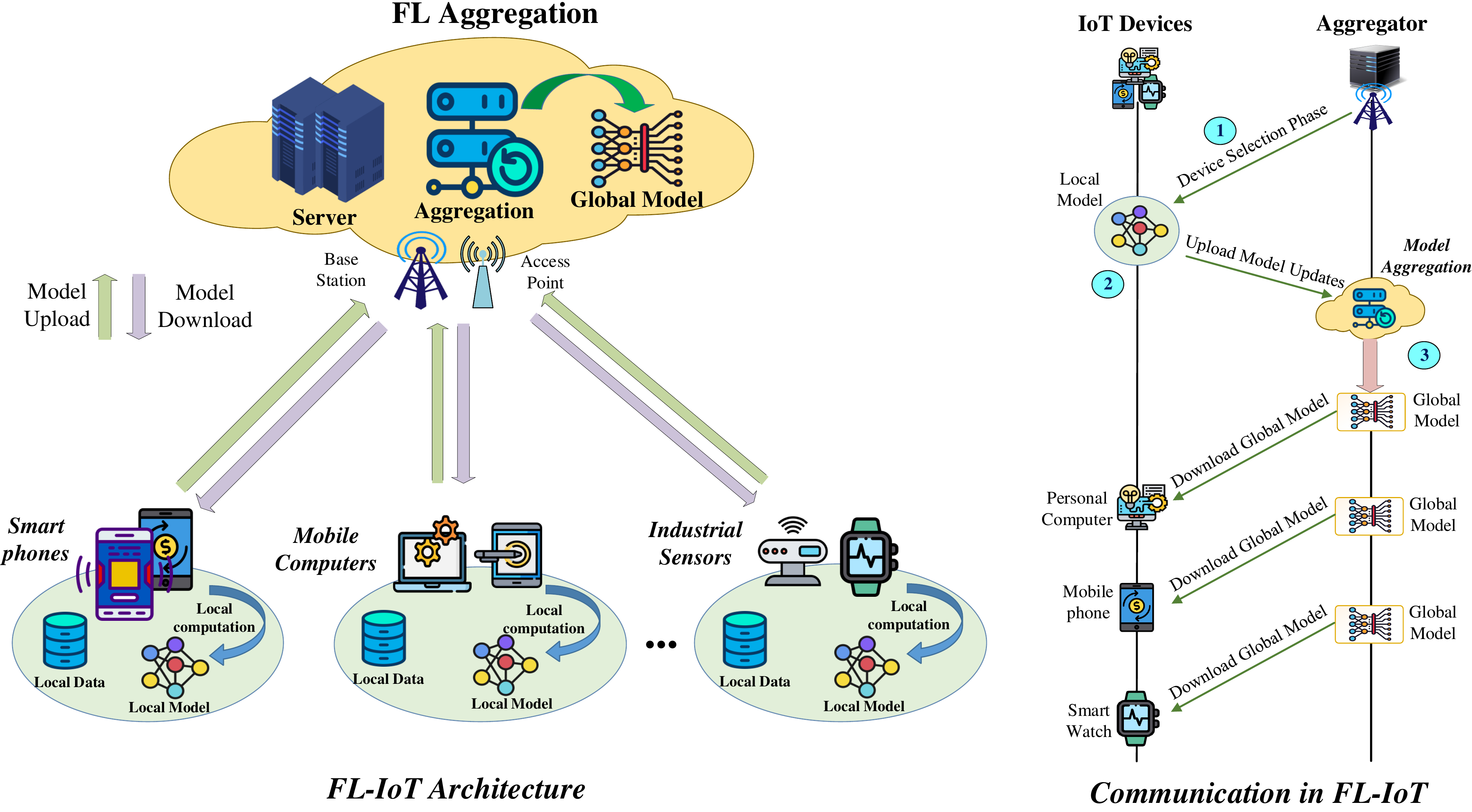}
	\caption{\textcolor{black}{The network architecture and communication process for FL-IoT. }}
	\label{Fig:FL_Concept}
	\vspace{-0.1in}
\end{figure*}
\begin{enumerate}
	\item \textit{System Initialization and Device Selection:} The aggregator chooses an IoT tasks such as human activity recognition and sets up learning parameters, e.g., learning rates and communication rounds. A subset of IoT devices to participate in the FL process is also selected. Several possible selection factors for this selection can be channel conditions and the importance of local updates of each IoT device \cite{13}. 
	\item \textit{Distributed Local Training and Updates:} After the training configuration, the server  initializes a new model, i.e., $ \textbf{w}_G^0$, and transmit it to the IoT clients to start the distributed training. Each client $k$ trains a local model using its own dataset $D_k$ and computes an update $ \textbf{w}_k$ {by minimizing a loss function} $\mathcal{F}( \textbf{w}_k)$:
	\begin{equation}
	\textbf{w}_k^* = \arg\min \mathcal{F}( \textbf{w}_k), k \in \mathcal{K}. 
	\end{equation}
	Here, the loss function can be different for different FL algorithms \cite{19}. For example, with a set of input-output pairs $\{x_i,y_i\}_{i=1}^K$, the loss function $\mathcal{F}$ of a linear regression FL model can be defined as: $\mathcal{F}( \textbf{w}_k) = \frac{1}{2}\left(x_i^T \textbf{w}_k -y_i\right)^2$. Then, each client $k$ uploads its computed update $\textbf{w}_k$ to the server for aggregation.
	\item \textit{Model Aggregation and Download:} After collecting all model updates from local clients, the server aggregates them and calculates a new version of global model as
	\begin{equation}
	\textbf{w}_G = \frac{1}{\sum_{k \in \mathcal{K}}|D_k|}\sum_{k=1}^{K}|D_k|\textbf{w}_k,
	\end{equation}
	by solving the following optimization problem: 
	\begin{equation}
	\begin{aligned}
	& (P1):   \underset{\textbf{w}_{k \in \mathcal{K}}}{\min }
	&& \frac{1}{K}\sum_{k=1}^{K}\mathcal{F}(\textbf{w}_k) \\
	& \text{subject to}
	&&(C1): \textbf{w}_1 =\textbf{w}_2=...=\textbf{w}_K=\textbf{w}_G. 
	\end{aligned}
	\end{equation}
	Here, the loss function $\mathcal{F}$ reflects the accuracy of the FL algorithm, e.g., the accuracy of an FL-based object classification task \cite{20}. Moreover, the constraint (C1) ensures that all clients and the server shares the same learning model over the FL task after each training round. {After the derivation of the model}, the server broadcasts the new global update $\textbf{w}_G$ to all clients for optimizing the local models in the next learning round. The FL process is iterated until the global loss function converges or a desired accuracy is achieved. 
\end{enumerate}
\subsubsection{FL Classifications}
Reviewing the recent advances of FL algorithms used in IoT, we here classify FL into two key dimensions, namely \textcolor{black}{data partitioning and networking structure \cite{10,11}.} 
\begin{figure*}
	\centering
	\includegraphics[width=0.98\linewidth]{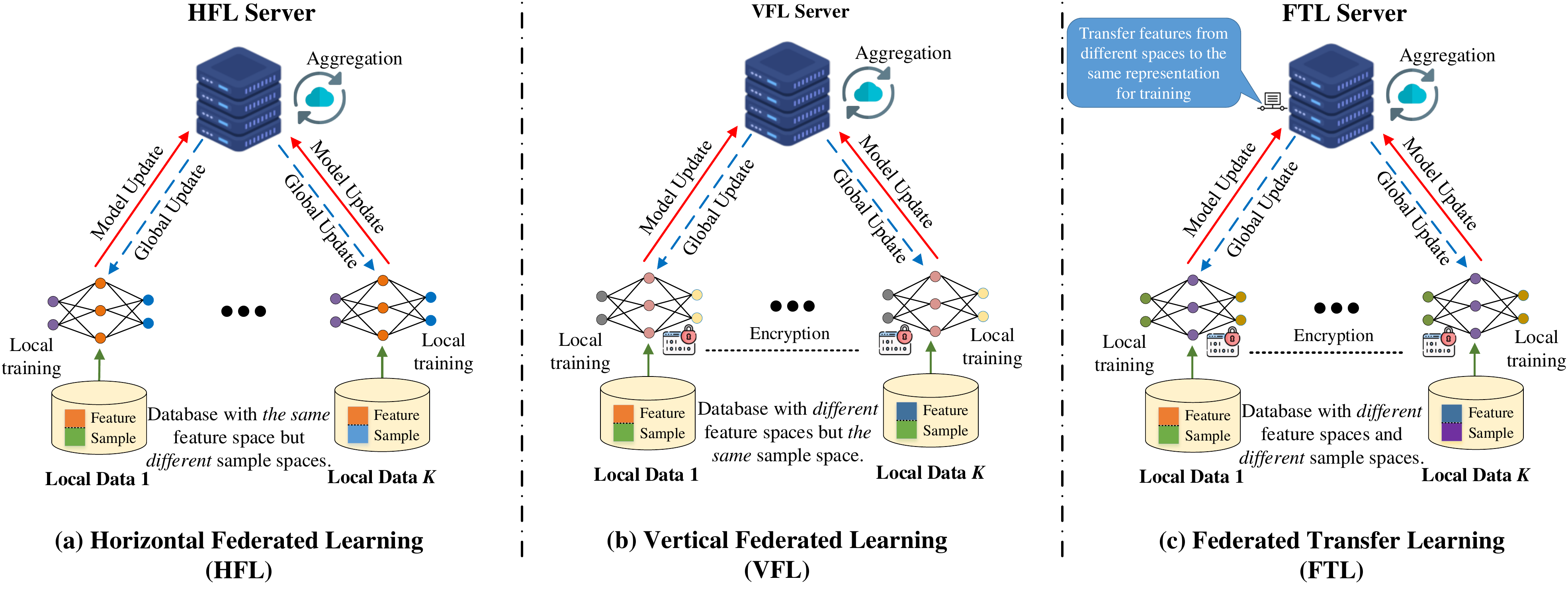}
	\caption{\textcolor{black}{Types of FL models with data partitioning.}}
	\label{Fig:FL_Classifications}
	\vspace{-0.1in}
\end{figure*}

\begin{figure*}
	\centering
	\includegraphics[width=0.98\linewidth]{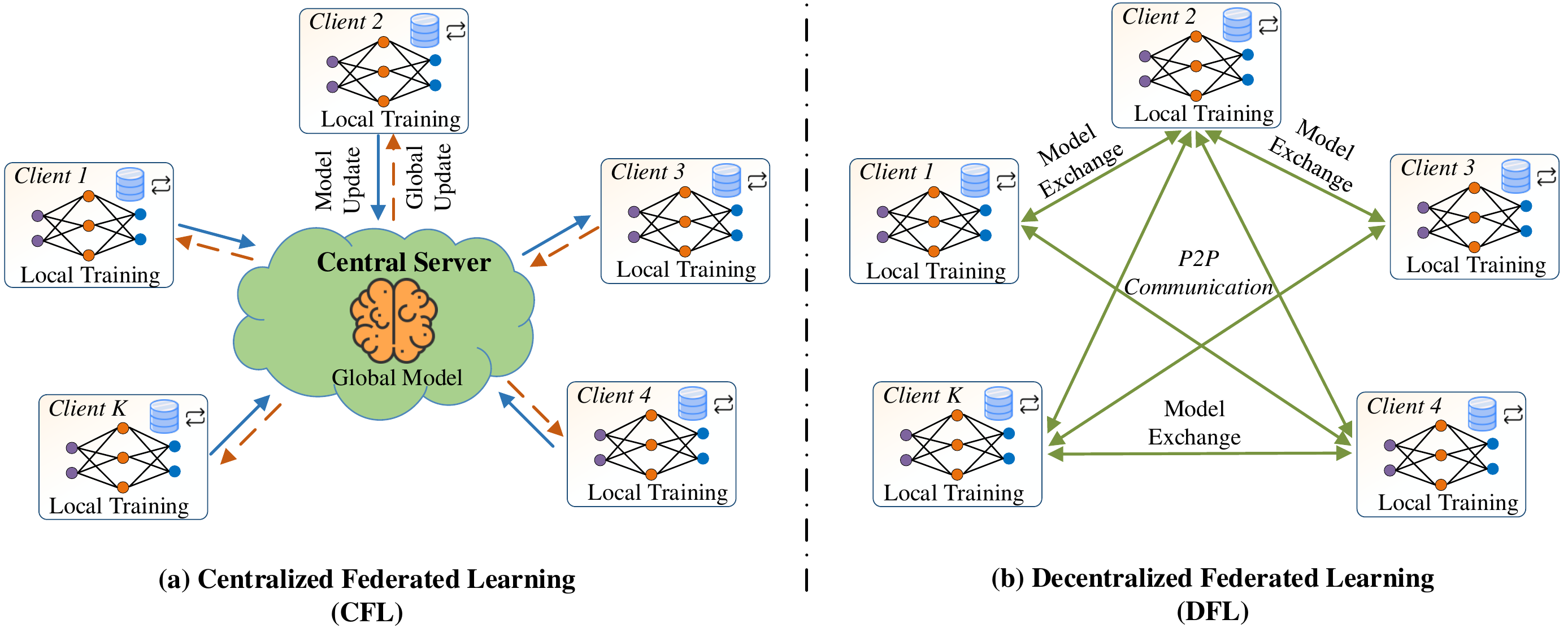}
	\caption{\textcolor{black}{Types of FL models with networking structure. }}
	\label{Fig:FL_Communication_Classification}
	\vspace{-0.1in}
\end{figure*}

- \textbf{\textcolor{black}{Data partitioning:}} Based on how training data are distributed over the sample and feature spaces, this category can be divided into three small classes, including horizontal FL, vertical FL, and federated transfer learning \cite{11} as summarized in Fig.~\ref{Fig:FL_Classifications}. 
\begin{itemize}
	\item \textit{Horizontal FL (HFL):} In HFL systems, all learning clients cooperatively train a global FL model using their local datasets with the same feature space but different sample space, as shown in Fig.~\ref{Fig:FL_Classifications}(a). Due to the same data feature, clients can use a same AI model (e.g., linear regression, SVM) for their local training. In a HFL system, each client locally trains its AI model to compute a local update. For improved security, the computed local update can be masked by using encryption or differential privacy techniques \cite{cp0}. Then, the server aggregates all local updates from clients and computes the new global update without the need for direct access to local data. Finally, the server sends back the global update to all clients for the next round of local learning. The above process iterates until the loss function converges or a desirable accuracy is achieved. In IoT applications, an example of HFL is Wake-word detection \cite{21}, e.g., voice assistants in a smart home. In this case, users speak the same sentence (feature space) with the different types of voice (sample space) on their smartphones and then the local speaking updates are averaged by a parameter server to create a global model for voice recognition.
	\item \textit{Vertical FL (VFL):} Different from HFL, VFL solves the shared AI model learning in a network of clients which have the same sample space with different data feature spaces \cite{22}, as shown in Fig.~\ref{Fig:FL_Classifications}(b). In VFL, an entity alignment approach is adopted to collect overlapped data samples of clients. These samples are combined to train a common AI model using encryption techniques. An example of VFL in IoT applications can be the shared learning model among entities in a smart city, e.g., e-commerce companies and a banking institution. In a smart city, an e-commerce company and a bank (different data feature) which serve city customers (same sample space) can join a VFL process to cooperatively train an AI model using their datasets, e.g., historic user payment at e-commerce companies and user account balance at the bank. Using this model, VFL can estimate the optimal personalized loans for all customers based on their online shopping behaviours. 
	\item \textit{Federated Transfer learning (FTL):} FTL \cite{23} aims to extend the sample space from the VFL architecture with more learning clients that have datasets with different sample space and different feature space, as shown in Fig.~\ref{Fig:FL_Classifications}(c). FTL transfers features from different feature spaces to the same representation that is used to train data aggregated from multiple clients. Also, to preserve data privacy and ensure security in the learning, encryption techniques such as random masks are also employed to encrypt gradient updates in the model update stage. Based on the combined updates from learning participants, the aggregation server can perform model learning to find the global update by minimizing a loss function \cite{24}. In IoT networks, FTL can be used in various domains, such as federated healthcare. For example, FTL can support disease diagnosis by collaborating different countries with multiple hospitals which have different patients (sample space) with different medication tests (feature space). In this way, FTL can enrich the shared AI model output for improving the accuracy of diagnosis. 
\end{itemize}
- \textbf{\textcolor{black}{Networking structure: }} From the networking perspective, this category can be divided into two small classes, including centralized FL and decentralized FL \cite{10} as summarized in Fig.~\ref{Fig:FL_Communication_Classification}. 

\begin{itemize}
		\item \textit{Centralized FL (CFL):} CFL is one of the most popular FL architectures used in FL-IoT systems. As shown in Fig.~\ref{Fig:FL_Communication_Classification}(a), a CFL system contains a central server and a set of clients to perform an FL model. In a single round of training, all clients participate in training a network model in parallel using their own datasets. Then, all the clients transmit the trained parameters to the central server which aggregates them by using a weighted averaging algorithm such as Federated Averaging (FedAvg) \cite{300}. Then, the computed global model is sent back to all clients for the next round of training. At the end of the training process, each client achieves a same global model along with its personalized model. In CFL, the server is regarded as the key component of the network for coordinating the aggregation and distributing the model updates to the clients to accomplish an FL task, such as object detection \cite{301}, while keeping training data secure and private.
		\item \textit{Decentralized FL (DFL):} Unlike CFL, DFL is a network topology without any central server to coordinate the training process. Instead, all clients are connected together in a peer-to-peer (P2P) manner to perform AI training, as shown in Fig.~\ref{Fig:FL_Communication_Classification}(b). In this way, in every communication round, the clients also perform local training based on their own dataset. Then, each client implements model aggregation using the model updates received from neighbor clients through the P2P communication to achieve a consensus on the global update \cite{302}. DFL is designed to fully or partially replace CFL when the communication with the server is not available or the network topology is highly scalable. Due to the contemporary features, DFL can be integrated with P2P-based communication technologies such as blockchain \cite{nguyenintegration2020} to build decentralized FL systems. Such that, the DFL clients can communicate via blockchain ledgers where model updates can be offloaded to the blockchain for secure model exchange and aggregation \cite{303}. In the following sections, some DFL systems based on blockchain will be explained in details. 
\end{itemize}

\subsection{Internet of Things}
With its ubiquitous sensing and computing capabilities, IoT is envisioned to connect a wide range of objects and things to the network, aiming to facilitate customer services and applications. To provide intelligence for IoT systems, AI techniques such as ML and DL have been widely adopted in IoT for enabling intelligent IoT systems \cite{25} thanks to their ability to discover knowledge of IoT data and get insights for realizing different smart applications, e.g., human activity classification, vehicular traffic control, weather prediction. In the following, we analyze two key aspects of IoT networks, namely IoT data analytics and intelligent service provision. 
\subsubsection{IoT Data Analytics}
In IoT networks, AI techniques can be integrated to build data analytic functions to process data collected from ubiquitous IoT devices such as sensors, actuators, smart phones, personal computers, and radio frequency identifications (RFIDs). ML/DL approaches such as neural networks have advanced computational models with multiple processing layers to learn the representations of data from different levels of abstraction, which helps extract useful features from natural data without requiring sophisticated feature engineering and tuning. These features make them attractive for handling different types of IoT data such as different modalities (image, time-series, video and text) or big data volume (data streaming from millions of sensing devices) \cite{25}. For example, a popular AI technique called Recurrent Neural Network (RNN) includes multiple links with neurons that are connected together to create a directed graph which allows RNNs to process data as temporal sequences of variable lengths in data processing tasks. The details of popular AI techniques used in data analytics can be found in our recent survey \cite{26}.	
\subsubsection{	Intelligent Service Provision}
Based on intelligent data analytics, AI can offer a number of IoT services. For example, RNN can be used to predict traffic flow by using a graph of a vehicular road network \cite{27}. That is, the topology of the road map is transformed into a spatio-temporal graph to create a structural RNN, which then can capture spatio-temporal features of the traffic speed data in time-series forms from sensors mounted on road segments. Another work in \cite{28} leverages a set of ML classifiers such as Bayes network, random forest, and support vector machine (SVM) to provide intelligent indoor localization based on data collected from capacitive sensors in room settings. Moreover, DL has been used in \cite{29} for human activity recognition. Here, an RNN model is built to extract deep features from Wi-Fi signal data via offline training that allows to remove redundancy information in raw data. Then, a Long-short Term Memory (LSTM) architecture is combined with RNN to provide further feature extraction for better detection accuracy with low computational complexity. Such representative use cases clearly demonstrate the great potential of AI techniques in IoT networks. With the help of FL, AI can achieve a much better level of scalability and privacy preservation by coordinating multiple devices to perform AI training at the edge of IoT networks while keeping data safe and secure at local devices. This also opens the door for newly emerging IoT services and applications which will be explored in this paper. 
\subsection{{Visions of the Use of FL in IoT}}

\begin{figure}
	\centering
	\includegraphics[width=0.98\linewidth]{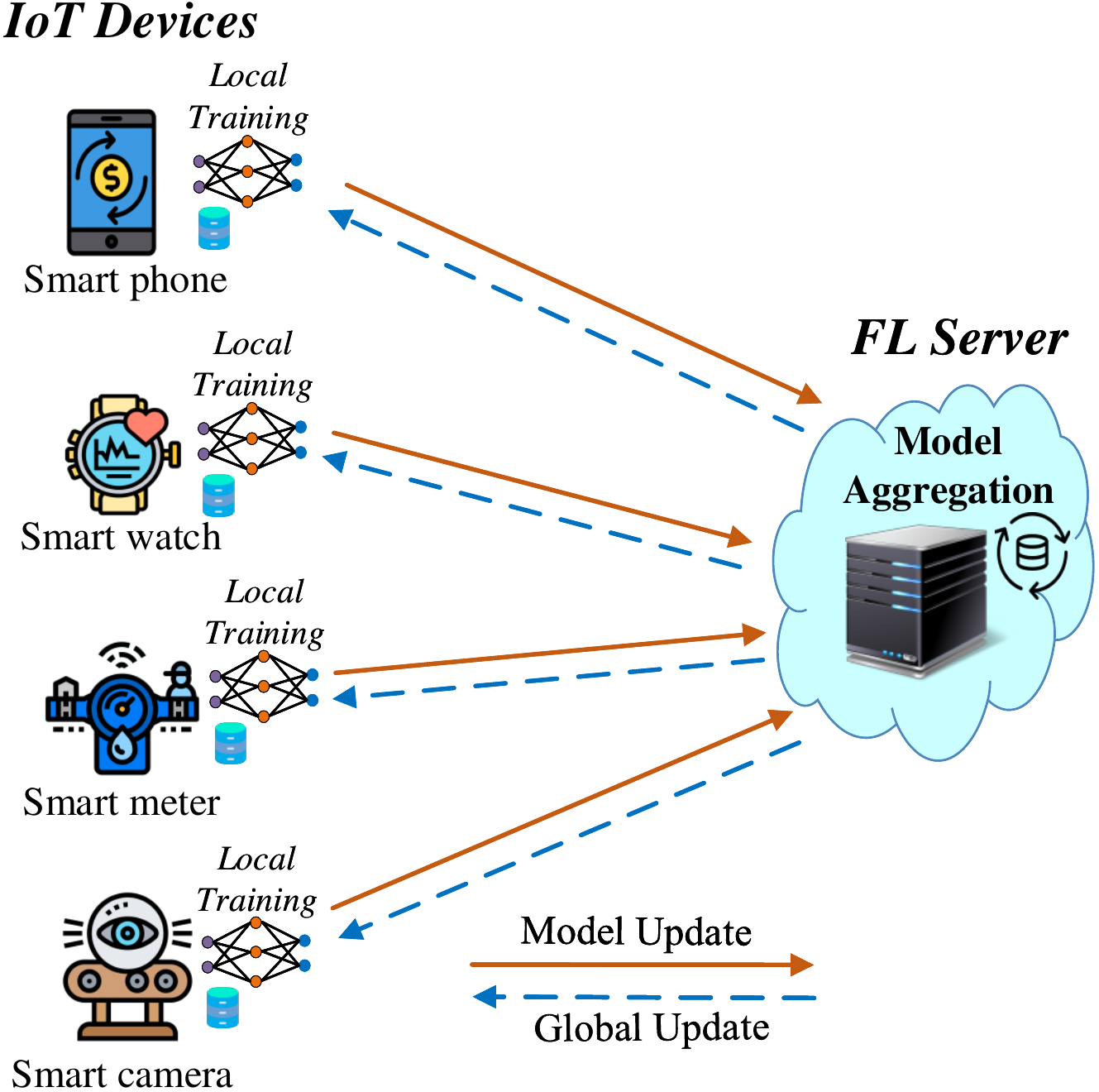}
	\caption{\textcolor{black}{A typical FL-IoT system.}  }
	\label{Fig:FL_IoT}
	\vspace{-0.1in}
\end{figure}

In traditional IoT systems \cite{25}, \cite{26}, AI functions are often placed on data centers or cloud servers, which is obviously not scalable to the exponential growth of IoT devices and the high data distribution of large-scale IoT networks. \textcolor{black}{Moreover, given a massive amount of ubiquitous IoT datasets in the big data era \cite{Cisco}}, it is infeasible to transmit a huge volume of data over complex environments to the data center for AI training. FL can provide more attractive features for enabling distributed intelligent IoT services and applications by using the computational capabilities of multiple IoT devices for data training, as shown in Fig.~\ref{Fig:FL_IoT}. This new architecture not only provides high quality of experiences for users in terms of low communication latency and privacy protection, but also benefits network operators such as channel bandwidth savings and more efficient computing resources of network servers for AI implementation. Indeed, FL has the great potential to transform current IoT systems, with many newly emerging services and applications. For example, the use of FL enables distributed data learning by leveraging computational capabilities of IoT devices to achieve an common objective, such as offloading latency minimization \cite{41}. Each mobile device acts as an FL client to train a local model and transmit the computed updates to an aggregator for overall model computation. The federation of mobile devices helps eliminate the need for a centralized data processing architecture, instead of performing data training locally using their own dataset without degrading learning performances and compromising user privacy values. FL is also promising to enable other IoT services such as attack detection or localization. For example, enabled by the {privacy enhancing features} of FL, federated attack detection and defense solutions can be realized using FL where each IoT device joins to run an AI model, such as a DNN, in order to train the threat model to fight against adversaries \cite{57}. The cooperation of multiple devices accelerates the learning process and improves learning accuracy while mitigating the risks of attack on the model learning. Moreover, FL is integrated with a centralized indoor localization model \cite{73} that relieves fingerprint collection workload and reduces network costs with privacy awareness, forming a decentralized indoor localization scheme by using the computational capability of distributed mobile devices. The usefulness of FL thus opens new opportunities for emerging localization services, such as localization in mobile indoor networks with global positioning system, mobile target tracking and navigation.

In addition to that, FL is able to provide new directions for enabling smart IoT applications, such as smart healthcare, smart transportation, and smart city. In fact, FL has the potential to support smart healthcare and reshape the current intelligent healthcare systems by proving AI functions for supporting healthcare services while {enhancing user privacy} and low latency with the cooperation of multiple entities such as health users and healthcare providers across medical institutions \cite{87}. Furthermore, FL has been introduced to bring AI functions to the network edge to empower smart transportation, involving a number of participants, such as vehicles, to collaboratively train globally shared AI models without the need for long data transmission and compromising user privacy. Some possible applications of FL in smart transportation can be vehicular traffic planning and vehicular resource management. FL is also a very useful solution to replace traditional centralized ML approaches in traffic prediction tasks \cite{112,115} by running ML models directly at the edge devices, e.g., vehicles, based on their datasets such as road geometry, traffic flow and weather. The use of massive data from multiple vehicles and {the large computational capability of all participant help} provide better traffic prediction outcomes, which cannot be achieved by using centralized ML techniques with less dataset and limited computation. FL has been exploited to provide distributed AI functions for decentralized smart city applications such as intelligent smart city data management \cite{140}. In this context, FL is helpful to structure data streams from ubiquitous IoT devices that work as FL clients for performing local learning without sharing their data to external third-parties. This would reshape the current forms of smart cities by providing newly exiting services such as smart urban communication, social economy sharing, social activity monitoring, and interconnection of global citizens.

In the following sections, we will present an extensive survey on the use of FL for IoT services and applications through different use cases. The roles and benefits of FL in IoT systems are discussed in details, and some key lessons obtained from the survey are also highlighted.

%% file: FL_Services.tex
\section{FL for IoT Services}
\label{Sec:FL_Services}
In this section, we present a state-of-the-art survey on the use of FL for IoT services, including IoT data sharing, data offloading and caching, attack detection, localization, mobile crowdsensing, and IoT privacy and security.  
\subsection{{FL Serving as an Alternative to IoT Data Sharing}}
Data sharing is one of the key services in IoT, aiming to transfer data over a shared network to serve end users in a specific application. {Instead of sharing the raw IoT data, FL offers an alternative of sharing learning results to enable intelligent IoT networks with low latency and privacy preservation.} The work in \cite{30} presents a collaborative data sharing model for industrial IoT applications where data owners and data requestors can achieve secure and fast data exchange among decentralized multiple parties. \textcolor{black}{Due to the resource constraints of IoT users, an FL scheme is designed that enables the estimation the types of data sharing requests with queries submitted by a requester to return the correctly computed results towards these queries for sharing.} Moreover, to protect the sharing process against external attacks, a blockchain \cite{31} is integrated with the FL architecture to build immutable data blocks that are controlled by all parties for transparency and improvement of data ownership without the need for any central authority. Once the sharing requests or data are recorded in the blockchain, they cannot be modified or changed which thus mitigate the risks of data leakage and improve network security accordingly. Simulation results from real-time datasets confirm the effectiveness of the proposed FL-based sharing scheme with high learning accuracy and improved security. In line with the discussion, the authors in \cite{32} develop a federated tensor mining framework in industrial IoT based on the FL concept to integrate multisource data for enabling tensor-based mining with security guarantees. More specifically, multiple factories cooperate to join the tensor mining by sharing their data, which have been encrypted using a homomorphic encryption technique, with a centralized server. Such that, the server only collects the ciphertext data and federates them into a tensor, while raw data are kept at local factories, which thus protects data privacy for the tensor mining. Although eavesdroppers can attack the centralized server to compromise the aggregated ciphertext and attackers can read ciphertext on communication channels, they cannot get the key for data decryption. By using two industrial alliances in Beijing and Shanghai with over 100 IoT nodes per factory for simulations, the proposed federated model can be improved by 24\% on mining accuracy compared to the privacy-preserving compressive sensing method, and provide high security degrees and efficient attack detection without performance degradation. 

\begin{figure*}
	\centering
	\includegraphics[width=0.98\linewidth]{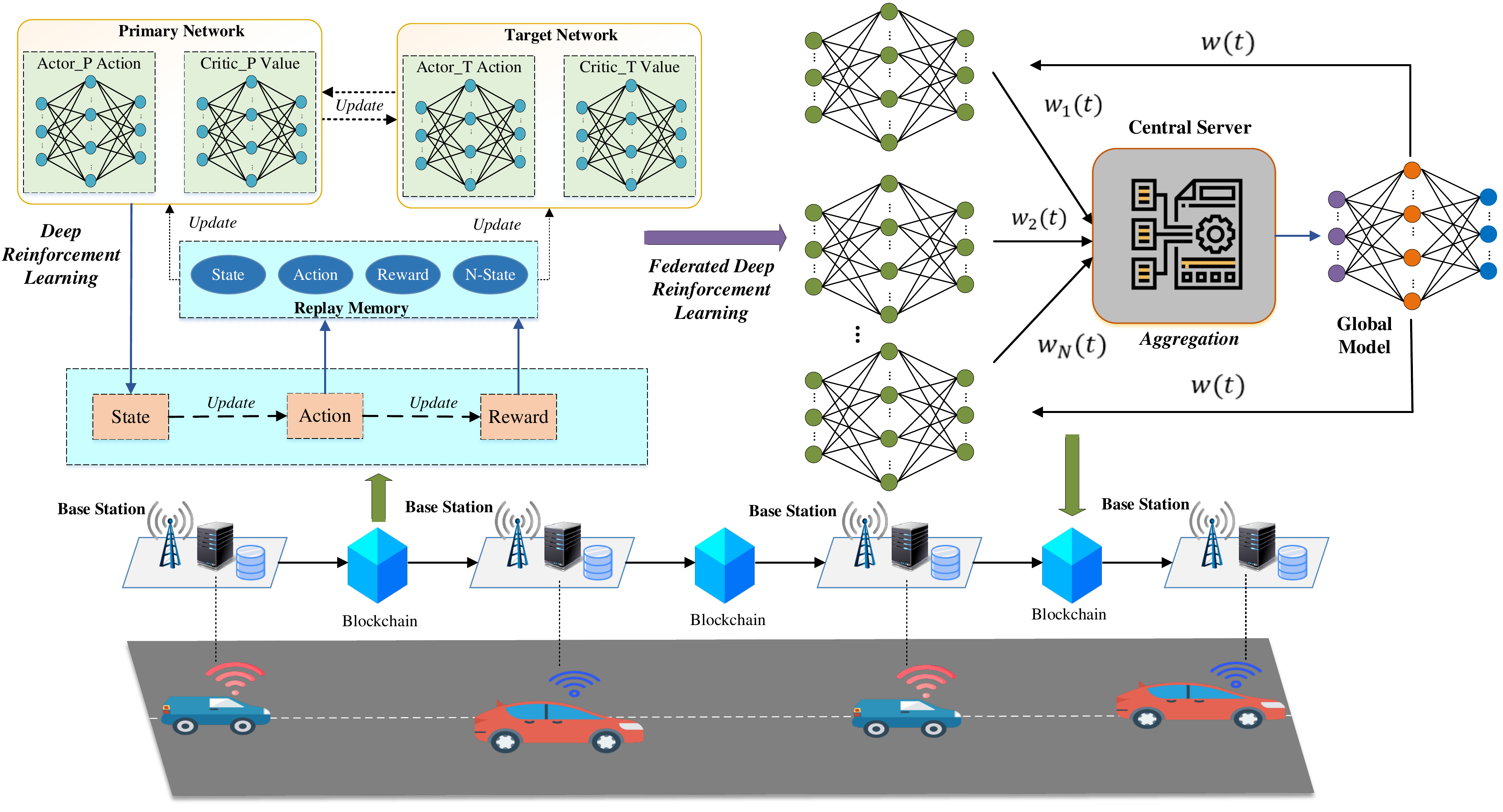}
	\caption{The architecture of federated learning-based data sharing for IoV with blockchain \cite{33}.  }
	\label{Fig:FL_DataSharing}
	\vspace{-0.1in}
\end{figure*}

FL has been applied to realize distributed data sharing in vehicular networks. The research in \cite{33} suggests an asynchronous federated data sharing framework for Internet of Vehicles (IoV) where each vehicle acts as an FL client to cooperatively share data with an aggregation server at a macro BS (MBS). Vehicles with different service demands such as traffic prediction or path selection can make a data sharing request to the MBS. \textcolor{black}{By performing a shared global model based on accumulated vehicular datasets, the MBS transforms the sharing process into a computing task to solve sharing requests of vehicles based on an actor-critic reinforcement learning framework. This approach learns to analyze the behaviors of participating nodes classified as bad or good nodes, aiming to support the intelligent data sharing decision by selecting the optimized participating nodes with sharing cost minimization.} To further improve security and reliability of the data sharing, blockchain has been used that aims to verify the model parameters and store them in immutable ledgers. The diagram for such a data sharing scheme based on FL and blocckhain is illustrated in Fig.~\ref{Fig:FL_DataSharing}.
Moreover, an asynchronous FL scheme for resource sharing in vehicular networks is also introduced in \cite{34}. Here, a local differential privacy mechanism that can preserve the privacy of local updates, is integrated into gradient descent training for enabling secure and robust FL sharing. Particularly, to avoid high communication cost and reduce security risks caused by the centralized FL architecture, a decentralized FL model is designed that allows to aggregate vehicles' model updates at distributed MBSs. Experiments from practical datasets show the promising results of the proposed decentralized FL scheme with better accuracy and privacy protection over traditional centralized FL approaches. An FL scheme is also considered in \cite{35} for knowledge sharing in vehicular networks with hierarchical blockchain. The proposed sharing architecture includes two main chain, i.e., ground chains and a top chain. To be clear, ground chains contain multiple vehicles as FL clients to implement local learning using their own hardware and road side units (RSUs) as decentralized FL aggregators that work in a blockchain network to securely collect transactions within their coverage. Meanwhile, the top chain maintains multiple RSUs that are responsible for performing FL model computation. The FL results are then appended into the block ledger for sharing among RSU and vehicles for security and traceability.

Furthermore, to facilitate the data sharing of non-independent and identically distributed (non-IID) among edge devices, a HFL scheme is proposed in \cite{36} for the shared learning between edge devices as participants and a cloud server as the aggregator. To deal with the issue of weight divergence caused by traditional FL, a federated swapping model is further developed based on a few shared data during the HFL that can mitigate the adverse impact of non-IID data. \textcolor{black}{Also, a semisupervised learning scheme is adopted to predict objects for video analysis applications among edge devices. By using real-world video data, the proposed FL scheme can achieve high accuracy in image classification, with an increase of 3.8\%, and the overall performance of object detection task is improved by 1.1\%, compared to the conventional FedAvg algorithm.} However, the reliance on a remote cloud for FL operation can result in long communication latency. For this problem, the authors in \cite{37} suggest a cost-efficient optimization framework that can coordinate edge devices and cloud {by minimizing the communication latency}. In this regard, the scheduling of shared data and admission control along with accuracy tuning are jointly optimized. The simulation results verify the feasibility of the proposed algorithm with reduced latency and {privacy enhancement} in various network settings. Similarly, the study in \cite{38} also considers an FL scheme in a mobile IoT network consisting of cloud servers and mobile devices as learning clients. The potential of FL in mobile cloud is investigated via a video recommendation system where each cloud collaboratively trains a local FL algorithm enabled by a dual-convolutional probabilistic matrix factorization model with user profile and textual information of videos as non-IID datasets. By using computed local updates, a federated recommendation algorithm is designed to enable information fusion among clouds with communication latency awareness. A drawback of this proposed scheme is the lack of security protection design for cloud-based learning in the FL process. To overcome this challenge, a secure data collaboration framework is studied in \cite{39} for federated data learning among multiple parties, including public data center and private data center, enabled by a blockchain ledger. The data usage event and FL updates are offloaded to the blockchain before transmitting to the central server for secure aggregation, while the data control of users is ensured. 
\subsection{{FL for the Optimization of IoT Data Offloading and Caching}}
\label{Subsection:Dataoffloading}
In addition to data sharing, FL has proved its strong ability to facilitate IoT data offloading and caching services in many applied domains.
\subsubsection{FL-based Optimization for IoT Data Offloading} 
Recently, the role of FL in IoT data offloading has been investigated. For example, an FL-based data offloading scheme for IoT networks is proposed in \cite{40} based on deep reinforcement learning (DRL). \textcolor{black}{Such that, multiple IoT devices act as DRL agents to build an intelligent offloading policy for offloading cost minimization with respect to different task probabilities via the FL concept.} This solution not only enhances data privacy due to the distribution of data learning in different agents but also mitigates the burden posed on the IoT network due to the centralized offloading architecture. Moreover, the cooperation of IoT devices under the guidance of the FL process is able to improve the overall training accuracy of the learning model. Simulations from different offloading task probabilities indicate an improvement in the offloading performance in terms of high system utility and low task transmission latency. The use of FL for edge data offloading is also investigated in \cite{41} where mobile devices collaboratively learn a shared offloading model in edge computing. The key objective is to maximize learning accuracy with respect to offloading latency and energy consumption constraints. Each FL client (or mobile device) acts as a local optimizer to solve a Lagrangian Dual problem formulated to achieve optimal offloading. Based on the proposed FL scheme, the offloading performance can be improved in terms of better accuracy compared to the heterogeneity-unaware equal task allocation approach \cite{42}. An FL-empowered task offloading problem with mobile edge computing (MEC) has been also analyzed in \cite{43}. \textcolor{black}{FL is combined with SVM to predict the user association for minimizing energy costs spent on task computation and transmission history. This can be done by using a learning federation approach with FL which allows each mobile user to learn an SVM model using its own dataset and then exchange the local updates for computing a global model based on the constraints of user association and task data size. As a result, the cooperation of users helps the BS to determine the best user association for offloading, aiming to reduce the energy consumption with high learning accuracy.}

In terms of video data offloading, the authors in \cite{44} suggest an FL-based solution for federated video analytics systems. An MEC server is built as a controller to control the resize rate by choosing different NNs to adjust the size of frames offloaded in the network. Then, edge devices and the MEC server join to learn an FL algorithm to guide the video offloading process across the edge network in a fashion the learning accuracy is ensured while the offloading latency is minimized. Meanwhile, an offloading scheme for vehicular networks with FL is recently considered in \cite{45}. Due to the user privacy concerns raised by the direct data offloading, FL has been used to build a cooperative learning architecture in which vehicles and RSUs can share a common learning model. This advanced approach potentially reduces learning costs with respect to agent selection and data sharing in task offloading. Recently, FL is leveraged in \cite{46} to characterize an offloading solution for data processing tasks in fog computing networks. By using FL, each mobile device can perform locally model learning for offloading optimization with respect to resource limitations and model accuracy without sharing raw data with the fog server for privacy protection. In particular, mobile devices can exchange their model updates together to minimize their model errors and adjust learning rates in a way the learning cost is optimized. By simulating with MNIST datasets consisting of thousands of images, the proposed FL-based offloading scheme can improve network resource utilization without sacrificing the learning accuracy of local models. 
\subsubsection{FL-based Optimization for IoT Data Caching}
\label{Subsection:DataCaching}
IoT data offloaded from mobile devices can be cached by edge servers \cite{48} where FL can play an important role in establishing intelligent caching policies, in order to cope with the explosive growth of mobile data in modern IoT networks. The use of FL can help overcome the challenges faced by traditional learning approaches in terms of high privacy concerns since data users may not trust the third-party server and thus hesitate to offload their private data for caching. \textcolor{black}{As shown in \cite{49}, FL is very useful to build proactive data caching schemes in edge computing without the need for direct access to user data to predict the most popular files for caching. By using the FL concept, mobile users can download the NN model from a cache entity, such as an edge server, for local training before sending back the model to the server for aggregation in an iterative manner.} This enables data file recommendation at the server based on the similarity of users and the files using latent features. Compared to baseline methods such as Oracle (Future caching demands are known in advance), random (random content selection) and Thompson Sampling (win and loss-based selection), the FL-enabled scheme achieves a better caching efficiency with increasing cache size and communication rounds. FL is also adopted in \cite{50} to train DRL agents to manage communication and computation usage in MEC systems for task offloading and content caching at MEC servers. By using FL, user equipments (UEs) do not need to offload their data to the MEC servers; instead, they train the DRL model at local devices and only submit the model parameters to the server. This learning solution aims for privacy protection and spectrum resource savings as well as ensures the robustness of the learning due to less impacts on unbalanced data and poor communication at any learning client. The role of FL in supporting DRL training for content replacement in data caching is also verified in \cite{51} in which UEs can cooperatively learn a shared model and keep raw data local, while the cloud at a BS can learn a global model by averaging local updates. Particularly, the content replacement can be formulated as a Markov decision process that is cooperatively solved by UEs via the FL process. 
\begin{figure*}
	\centering
	\includegraphics[width=0.98\linewidth]{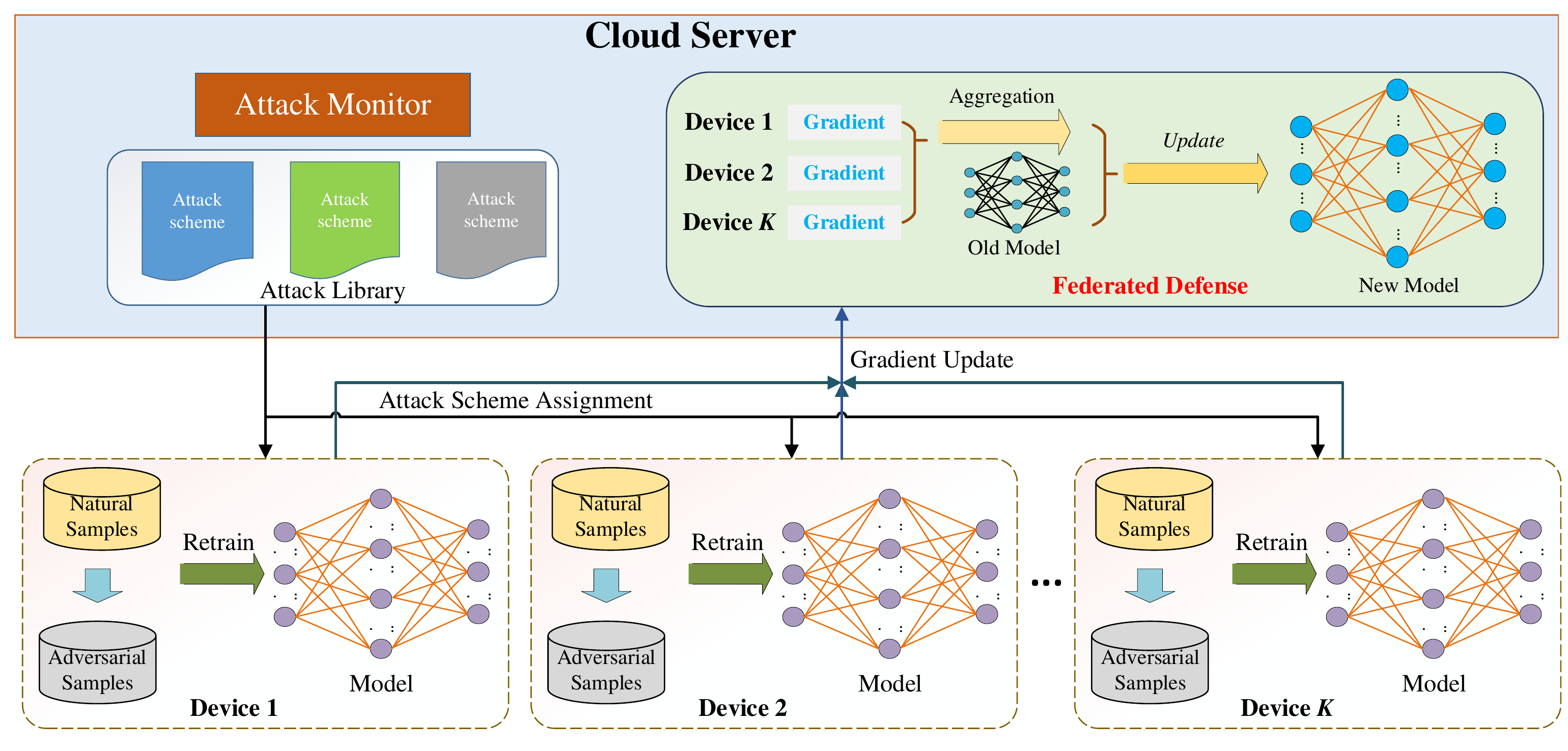}
	\caption{Federated attack detection and defense in FL-based IoT networks. }
	\label{Fig:FL_AttackDetection}
	\vspace{-0.1in}
\end{figure*}

To estimate the content caching placement at terrestrial BSs in UAV-based 6G networks, a 2-stage FL algorithm is designed in \cite{52} among mobile users, UAVs/BSs, and a heterogeneous computing platform. \textcolor{black}{In the FL architectures, each mobile user holds a deep neural network (DNN) model that contains shallow and dense layers. The parameters of shallow layers help learn the general features of content access, while the parameters of dense layers support the learning of specific content features and user context information. As a result, the global updates at the BSs can achieve a better insight of data traffic flow for improving caching performances.} Two real-word datasets including 100,000 ratings on 1682 movies from 943 users and over a million ratings on 3883 movies from 6040 users are employed to evaluate the FL algorithm, showing a better caching efficiency with improved learning accuracy. Recently, FL is integrated with blockchain to build secure learning schemes for content caching in edge computing \cite{53}. Smart contracts \cite{105}, a kind of self-executing program running on the blockchain, are employed to perform access verification and credit investigation during the content offloading and caching at the edge servers. Then, IoT devices participate in the local training using their own datasets to learn the features of users and files, and share the gradient updates to the edge server for popular file estimation, aiming to improve the overall cache hit rate. By compressing gradients at the local devices, the proposed scheme can further reduce communication overhead during the FL process. Simulations from MovieLens datasets confirm a significant improvement of caching efficiency over the traditional FL-based caching algorithms. Meanwhile, the authors in \cite{54}, \cite{55} pay attention to content popularity prediction with FL. In fact, the popularity prediction  of proactive edge caching at BSs can mitigate the traffic load and enhance user experience, by selecting the most popular and important contents in the offloaded data package. FL comes as a viable solution to estimate the file popularity without disclosing sensitive user information and user preference. In this context, a NN can be used to train the preference-weighted file popularity at local devices using historic request records, aiming to enhance the cache hit rate. By using FL, local updates can be averaged at a global edge server to predict popular files to be cached to balance the data usage experience of all pariticipants. 
\subsection{{FL for IoT Attack Detection}}
\label{Subsection:AttackDetection}
The popularity of IoT applications and services has become the main targets of malicious adversaries that can attack AI/ML models integrated in IoT networks, by modifying data inputs or changing learning network weights which can lead to erroneous  predicted outputs \cite{62}. Many solutions have been proposed to cope with attacks in IoT such as ensemble diversity or adversarial training \cite{56}, but they are mostly applied to a specific type of attack and not scale well to distributed IoT networks.  FL has emerged as a strong alternative to provided distributed intelligence for IoT systems with the ability to detect a wide range of attacks and support network defense solutions \cite{57}. \textcolor{black}{Enabled by the privacy enhancement feature of FL, a federated attack detection and defense solution is built in a fashion that each industrial IoT device joins to run a DNN model locally, in order to retrain the threat model to fight against adversaries.} In this process (as shown in Fig.~\ref{Fig:FL_AttackDetection}), each IoT device first produces adversarial samples to create a retraining set; then, the local trained updates are offloaded to a cloud server for synchronization. Finally, the cloud server computes a global model and then sends back to the local devices for the next round of learning. \textcolor{black}{The cooperation of multiple devices accelerates the learning process and improves the detection of adversaries while mitigating the risks of attack on the model learning.} In such a distributed learning environment, to ensure robust and safe FL operations, developing attack detection inside the FL architecture is essentially important. An attack detection approach is proposed in \cite{58} for robust and safe FL process. To be clear, a stochastic dynamical system is designed to evaluate the aggregated parameters in each FL round by using derive explicit conditions that prevent attacks from interfering the FL process at the aggregator. This work motivates more innovative solutions for attack detection and prevention for the FL, such as in \cite{59}. Here, a spectral anomaly detection mechanism is built at the global server to identify abnormal updates by verifying their low-dimensional embeddings to remove noise and potentally malicious samples while remaining useful data features. In practical leaning scenarios, clients such as IoT devices may be malicious ones and be curious about the model updates of other users at the global server. In this regard, the recognition and detection of malicious nodes are of paramount importance for safe FL \cite{60}. A solution using a pre-trained anomaly detection model is necessary to identify abnormal behaviors of clients in each communication round, by producing the surrogates of the local model weight updates at the global server. As a result, malicious clients and attacks can be detected and eliminated from the FL process while the federated model is preserved \cite{61}. \textcolor{black}{In addition to that, FL models can also be used to automatically detect adversarial clients in wireless networks, such as IoTs \cite{63}. Particularly, a dynamic linguistic quantifier is designed at the aggregation server to evaluate the weight contribution of each client in order to filter out adversarial clients and detect attacks such as model update poisoning attacks, data poisoning attacks, and evasion attacks.}

Moreover, FL also facilitates the detection of compromised IoT devices in federated IoT networks. In fact, with the sophistication of attacks and threats, it is challenging to detect them with current solutions \cite{65} that often recognize attacks by making deviations from normal behaviour profiles of users, which suffers from high false alarm rate and long detection delay. Moreover, IoT data are highly distributed over the large-scale network and each IoT device often has low computing power to run a detection algorithm. To overcome these challenges, FL comes as a natural solution to perform attack detection algorithms for distributed IoT networks \cite{64}. Each IoT device can submit a detection profile obtained by local training to a security gateway where local updates are aggregated to build a common detection model for the IoT network. The involvement of a large number of IoT devices with diverse features and massive datasets enhances the learning accuracy for better attack detection efficiency. Experiments with Mirai malware on a testbed confirm a high detection rate with 95.6\% with low learning time, compared to the centralized learning approach at an IoT server. Another FL-based attack detection scheme is also considered in \cite{66} where smart filters are built at IoT gateways to identity and prevent cyberattacks in {industry 4.0. More specifically,} each filter is built based on a DNN that trains local datasets collected from its subnetwork such as a smart farming or an energy plant. The trained model updates are then averaged at a central server to perform a detection algorithm. By integrating massive updates from multiple IoT gateway, the server can generate a high learning accuracy rate without compromising user privacy and saving network channel spectrum resources. FL has been also used to support federated intrusion detection in IoT \cite{67} for replacing traditionally centralized learning-based approaches. An important issue of designing intrusion detection systems is quick detection while user information is protected. FL can meet this requirement by distributing AI/ML models to the local devices and using the computational capability of all clients such as routers for building a strong attack defense mechanism for better detection rate. To improve the scalability of intrusion detection, a line-speed and scalable FL approach is introduced in \cite{68}. Each IoT device runs a binarized NN for packet classification at the line speed of nearby switches in a scalable manner, while preserving privacy of network traces. \textcolor{black}{Interestingly, FL can be deployed on mobile devices such as Android phones for malware detection \cite{69}. A safe semi-supervised ML algorithm is integrated at each Android phone that cooperatively contacts with other phones to build a shared malware classification algorithm at a data server. The results obtained from simulations show a high accuracy of malware app detection without false alarms.}
\subsection{{FL for IoT Localization}}
\label{Subsection:IoTLocalization}
The proliferation of smart devices enables a wide range of location-based services which often rely on localization systems. Wireless signals and channel state information (CSI) can be employed to determine the preferred reference locations of the target object. Traditional AI/ML-based localization systems often suffer from the lack of robustness due to the dynamics of mobile environments and privacy leakage bottlenecks caused by the centralized data processing architecture for object localization \cite{70}. FL can provide interesting solutions for enabling efficient and privacy-preserved localization services. As an example, FL is used in \cite{71} to build a privacy-preserving indoor localization model in residential building settings. \textcolor{black}{Mobile users can build their AI models locally by using received signal strength measurements from beacons with labelled locations, while the central server builds a global multilayer perceptron model from local updates for accurate localization estimation.} By using available RSS data from UJIIndoorLoc database \cite{72}, a simulation is performed for the proposed FL scheme, showing a significant improvement in the localization accuracy while privacy is preserved in different indoor settings. The potential of FL in indoor localization services is also verified in \cite{73} to solve the issues of repetitive task learning and privacy leakage risks due to the reliance of a central AI server. In the proposed architecture, FL is integrated with a centralized indoor localization model that relieves fingerprint collection workload and reduce network costs with privacy awareness, forming a decentralized indoor localization scheme by using the computational capability of distributed mobile devices. Each device runs a deep learning model using labeled fingerprint data and unlabeled crowdsourced data and then exchanges the computed updates with a central server for building a global statistical localization model. The use of FL clearly shows a better performance in terms of lower communication cost, privacy protection, and robustness in various fingerprint data distribution settings. 

Another work in \cite{74} also leverages FL to build a federated localization scheme for WiFi networks. By using WiFi signals that represents predefined reference locations, mobile devices can build their local fingerprints to run a DNN model followed by a deep autoencoder for noise removal. Then, the local weights are combined by a central server to generate a general model. Besides, to protect the communication channel during the offloading phase, a homomorphic encryption technique is adopted to encrypt local updates. An experiment is conduced in a laboratory corridor setting, indicating a high localization estimation accuracy with high security. \textcolor{black}{Similarly, a Federated Localization (FedLoc) framework is considered in \cite{75} for providing accurate localization services in IoT networks. By cooperating multiple users in the model learning using local fingerprint data, FL minimizes the bias of location estimation with privacy protections.} The usefulness of FL opens new opportunities for emerging localization services, such as localization in mobile indoor networks with global positioning system (GPS), mobile target tracking and navigation based on inertial sensors, and wireless traffic prediction with BSs. 

\subsection{{FL for IoT Mobile Crowdsensing}}
With the development of IoT, mobile crowdsensing is designed to take advantage of pervasive mobile devices for sensing and collecting data from physical environments to serve end users. In intelligent mobile crowdsensing systems, traditional AI/ML architectures have been deployed for model training and processing \cite{76}. However, these traditional crowdsensing solutions usually require direct access to user data, which leaves a chance for privacy leakage. Moreover, the use of a central server to handle all sensing data is not a scalable solution, making it hard to cope with the massive volume of data from ubiquitous IoT devices. FL would be a very promising tool to accelerate the learning and training for crowdsensing models. \textcolor{black}{In fact, FL has been used in recent works with encouraging results. {For example, FL is used to support automatic content suggestions for on-device keyboards \cite{171}, by sensing and suggesting relevant contents, e.g., search queries, based on input texts from a network of smartphones. The feasibility of FL has been tested in Gboard on Android phones where each phone runs a query suggestion model based on local on-device information and collaborates with a cloud server.} By using a SGD-based optimization model, the cloud can train the text prediction model to generate a suggested query when a user types on the phone.} The study in \cite{77} shows an FL-based mobile crowdsensing scheme, with a focus on privacy-preserving XGBoost training with the cooperation of multiple mobile users. A secure gradient aggregation algorithm is designed by integrating homomorphic encryption with secret sharing, which prevents the central server from guessing decryption result before operating aggregation. Simulations MNIST datasets reveal very competitive results, with high accuracy rate (over 98\%), and a reduction of 23.9\% runtime and 33.3\% communication latency for gradient aggregation. Similarly, a security mechanism is proposed in \cite{79} along with FL for secure XGBoost training in mobile crowdsensing with cloud computing. The FL model takes three main entities into account building the federated system, including users, edge servers and cloud. Here, users run classification and regression tree models at local devices and then offload the computed updates to the cloud for averaging protected by a security protocol built on the edge layer. Given a crowdsourcing/crowdsensing system, the work in \cite{80} focuses on designing an incentive mechanism \cite{83} by analyzing the interactions between the participating clients and the aggregator at an edge server to minimize learning costs. To be clear, each client selects a learning strategy for solving its local sub-problem to ensure desired accuracy with lowest participation costs, while the central server builds a utility function by averaging local updates to offer reward to the clients. This incentive process is modelled by a two-stage Stackelberg game that can outperform the heuristic approach in terms of a utility gain improvement by 22\% for different system settings. The game theoretic approach is also adopted in \cite{81} for the FL-based crowdsensing in IoT networks. MEC servers, cloud and IoT sensors cooperatively join to build a shared learning model in a fashion the utility of MEC operators is maximized by considering a tradeoff between the revenue and energy cost via a Stackelberg equilibrium formulation. 

\begin{figure}
	\centering
	\includegraphics[width=0.97\linewidth]{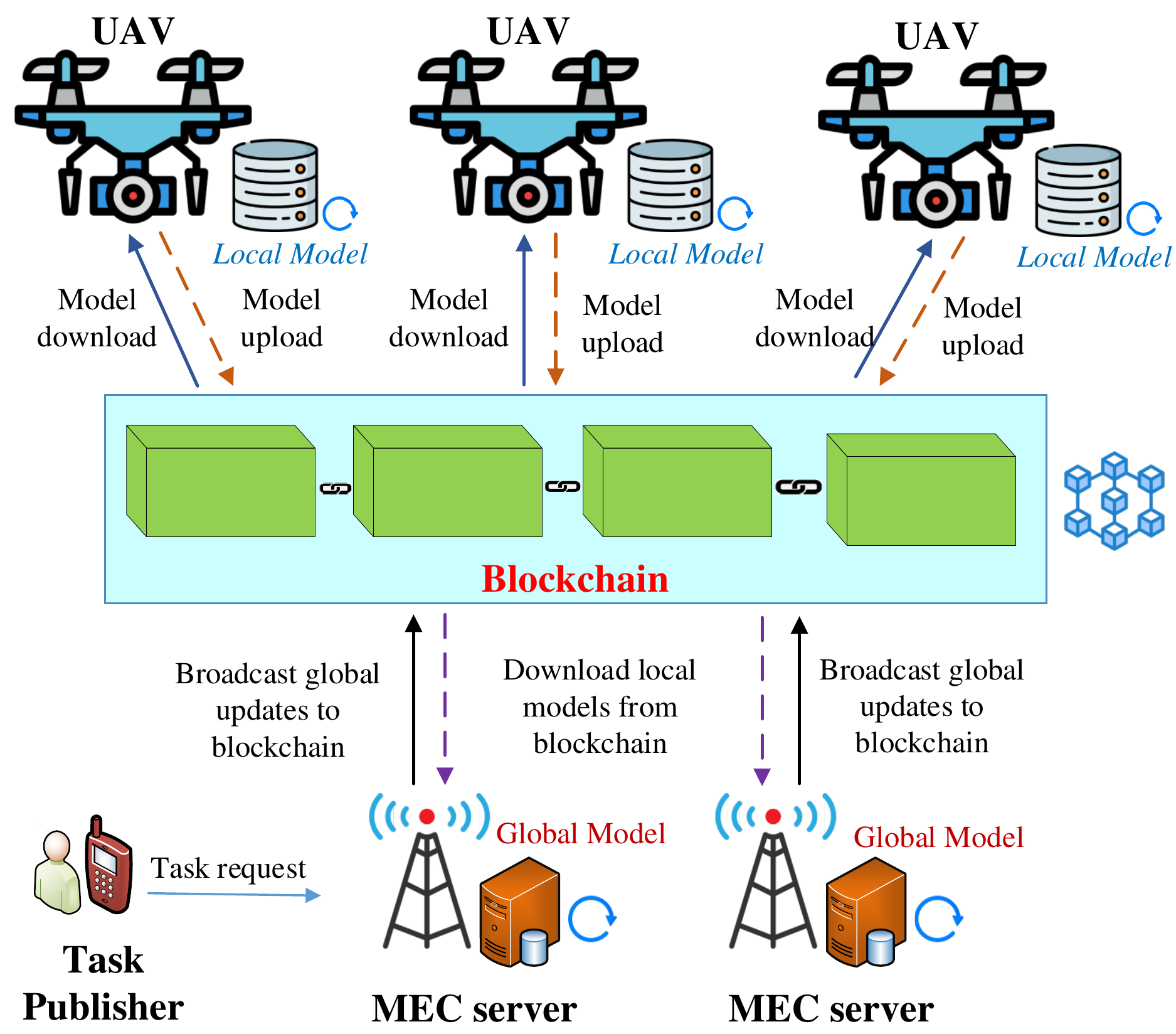}
	\caption{Federated learning for blockchain-based crowdsensing in UAV networks.   }
	\label{Fig:FL_Crowdsensing_UAV}
	\vspace{-0.1in}
\end{figure}

In traditional FL systems, the reliance on a central server for aggregation can result in high communication overhead and less robustness due to the heterogeneity of different IoT devices. The authors in \cite{82} solve these problems by providing a distributed FL framework in a sensing platform for distributed decision making and learning among IoT devices. To mitigate latency incurred by the central server communication, an infusing redundancy solution is considered to speed up distributed stochastic gradient descent calculation based on a distributed of datasets. In this way, each IoT node only needs to compute the model based on its mini-batch instead of the whole datasets, which will reduce computation latency accordingly. Recently, FL is also employed to build a secure UAV-based crowdsensing approach \cite{84}, as shown in Fig.~\ref{Fig:FL_Crowdsensing_UAV}. Instead of relying on a central aggregator, blockchain is introduced to decentralize the learning process by integrating UAVs with task publishers with a blockchain ledger that make the data training and contribution verification among UAVs secure and transparent. To further improve privacy for the local learning updates, a local differential privacy technique is adopted in communication rounds with aggregate accuracy guarantees. Based on that, an incentive mechanism is also built to evaluate the efficiency of the interactions between publishers and UAVs. Experiment results from a MNIST dataset including 60,000 training samples and 10,000 test samples with convolutional neural networks (CNNs) as the learning model at UAVs show a high utility of UAVs and low aggregation error along with low convergence latency. 

\subsection{\textcolor{black}{FL-based Techniques for Privacy and Security in IoT Services and Networks}}
\textcolor{black}{In IoT networks, security and privacy remain huge issues for IoT devices, which introduce a whole new degree of attacks and privacy concerns for consumers. This is because these devices not only collect personal information but can also monitor user activities. Many AI/ML algorithms have been adopted to address security and privacy issues, by their ability to classify and detect threats and privacy bottlenecks in IoT networks. However, these traditional solutions also have several limitations, including the need for centralized IoT data collection and user privacy exposure due to public data sharing. FL appears as an attractive approach for enabling intelligent privacy and security services in IoT networks. For example, FL has been used for data privacy preservation in vehicular IoT networks \cite{408} where two-phase mitigating scheme is proposed for intelligent data transformation and collaborative data leakage detection. Different from the existing schemes, the proposed vehicular FL solution allows participants (e.g., vehicles) to train models locally with their own data without a centralized curator, which contributes significantly to protecting their data privacy. Moreover, a joint mapping approach is performed over multiple vehicles to ensure the utility of data after the transformation, allowing for mapping raw data of multiple parties into learned data models. The learned model contains valid information to be further used in tasks such as resource allocation without revealing their raw data, which further enhances privacy protection. Similarly, to avoid the privacy threat to vehicular IoT networks, a new approach is suggested in \cite{409} using FL combined with local differential privacy which aims for perturbing gradients generated by vehicles while not compromising the utility of gradients. This would prevent attackers from deducing original data even though they obtain perturbed gradients. As a result, the FL server gathers and averages users' submitted perturbed gradients to obtain the average d result to update the global model's parameters without privacy concerns. In FL-IoT systems, a centralized entity is often used to perform model aggregation which introduces single-point failures and privacy concerns due to the curious server. Blockchain \cite{nguyenintegration2020} can be used to address this issue, aiming to replace the centralized aggregator in the traditional FL system \cite{410}. A consortium blockchain is used  to store FL model updatess permanently. More specifically, a manufacturer first uploads an initial model to the blockchain so that customers can send requests to obtain that model. After training models locally, customers upload their locally trained models to the blockchain and a hash will be sent to the blockchain as a transaction. The hash can be used to retrieve the actual data from blockchain storage. The leader and miners are responsible for confirming transactions and calculating the averaged model parameters to obtain a global model. To protect privacy of customers and improve the test accuracy, the scheme also enforces differential privacy on the extracted features for providing a new level of privacy for FL training. }

\textcolor{black}{Moreover, FL has been applied to protect security in IoT services and networks. For instance, the work in \cite{411} considers an NN-based FL scheme for network anomaly detection such that the participants do not share their training data to a third party, which can prevent the training data from being exploited by attackers. A multi-task learning method is proposed by using a distributed DNN which can perform the network anomaly detection task, traffic recognition task, and traffic classification task simultaneously. Experimental results demonstrate the effectiveness of FL in detecting network anomalies, achieving an accuracy of 97.81\% and closeness to the centralized scheme. In the current FL-IoT systems, clients are autonomous in that their behaviors are not fully governed by the aggregator. As a result, a client may intentionally or unintentionally deviate from the prescribed course of federated model training, which leads to abnormal behaviors, such as turning into a malicious attacker or a malfunctioning client. An FL-based approach is studied in \cite{412} where security is analyzed, aiming to detect anomalous clients at the server side by using low-dimensional surrogates of model weight vectors for anomaly detection. Recently, blockchain is also integrated into FL to solve security issues in fog-based IoT networks \cite{413}. With hybrid identity generation, comprehensive verification, and access control offered by decentralized blockchain, FL model updates are transmitted safely over the distributed fog network in an immutable manner. Also, the mining process performed by the miners allows for robust authentication on user access during the training against malicious threats and learning data breaches. }
	
In summary, we list FL-IoT services in the taxonomy Table~\ref{Table:FL_Services} and Table~\ref{Table:FL_Services_next} to summarize the technical aspects as well as the key contributions and limitations of each reference work.

\begin{table*}
	\centering
	\caption{Taxonomy of FL-IoT services. }
	{\color{black}
		\label{Table:FL_Services}
		\begin{tabular}{|P{0.5cm}|P{0.5cm}|P{1.5cm}|P{0.5cm}|P{01cm}|P{1.1cm}|P{1.3cm}|P{4cm}|P{4cm}|}
			\hline
			\textbf{Issue}& 	
			\textbf{Ref.} &	
			\textbf{Use case}&	
			\textbf{FL type}& 
			\textcolor{black}{\textbf{ML type}}& 
			\textbf{FL clients}& 	
			\textbf{FL aggregator}&
			\textbf{Key contributions}&
			\textbf{Limitations}
			\\
			\hline
			\parbox[t]{1.5cm}{\multirow{21}{*}{\rotatebox[origin=c]{90}{FL for IoT Data Sharing}}} & 
			\cite{30}&	Industrial IoT data sharing&	CFL&NN&	IoT users&	Data center&	A privacy-preserved data sharing scheme for industrial IoT with FL and blockhain. &	Communication cost for such a CFL is high. 
			\\ \cline{2-9}&
			\cite{32}&	Industrial IoT data sharing&	CFL&-&	Factories &	Data center&	A secure FL-based data sharing scheme for tensor mining in factories. &	The latency performance for the federated mining has not been evaluated. 
			\\ \cline{2-9}&
			\cite{33}&	Vehicular data sharing&	DFL&DRL&	Vehicles&	MBSs&	A secure asynchronous FL-based data sharing model for IoV. &	Security performance has not been investigated. 
			\\ \cline{2-9}&
			\cite{34}&	Vehicular data sharing&	DFL&Decision tree &	Vehicles &	MBSs&	A differential privacy-based FL model for secure vehicular data sharing. &	Numerical evaluations of security for FL-based sharing have not been implemented.
			\\ \cline{2-9}&
			\cite{35}&	Vehicular knowledge sharing&	DFL&CNN&	Vehicles& 	RSUs&	A hierarchical blockchain-based FL scheme for vehicular knowledge sharing. &	The impact of bloclchain consensus on FL sharing has not been analyzed. 
			\\ \cline{2-9}&
			\cite{36}&	IoT video data sharing&	HFL&CNN&	Edge devices&	Cloud server&	A sharing scheme for non-IID data among edge devices and cloud. &	The latency for cloud communication is high.
			\\ \cline{2-9}&
			\cite{38}&	IoT video data sharing&	CFL&CNN&	Cloud&	Cloud&	An FL scheme for video recommunication and sharing in mobile cloud. &	Security risks from cloud-based learning has not been investigated. 
			\\ \cline{2-9}
			\hline
			
			\parbox[t]{1.5cm}{\multirow{30}{*}{\rotatebox[origin=c]{90}{FL for IoT Data Offloading and Caching}}} & 
			\cite{40}&	IoT data offloading&	HFL&DRL&	Mobile devices&	Edge server&	An FL scheme for cooperative IoT data task offloading. &	Learning accuracy of FL-based offloading has not been analyzed.
			\\ \cline{2-9}&
			\cite{41}&	Edge data offloading&	VFL&-&	Mobile devices&	Edge server&	An FL framework for federated data offloading in mobile edge networks. &	The proposed FL scheme is simple and not scalable to multi-edge networks. 
			\\ \cline{2-9}&
			\cite{43}&	Edge data offloading&	HFL&SVM&	Mobile devices&	BS&	An FL-empowered task offloading in MEC. &	The feasibility of the proposed model in practical settings has not been investigated.
			\\ \cline{2-9}& 
			\cite{44}&	Video data offloading&	HFL&DNN&	Mobile devices&	MEC server&	An video data offloading scheme with FL in edge networks. &	The proposed offloading model is only applied to a specific application. 
			\\ \cline{2-9}&
			\cite{45}&	Vehicular data offloading&	DFL&- & 	Vehicles&	RSUs&	A federated task offloading in vehicular networks with FL. &	Data privacy should be considered in the task offloading.
			\\ \cline{2-9}&
			\cite{46}&	Fog data offloading&	HFL&CNN &	Mobile devices&	Fog server&	A data processing task offloading scheme with FL. &	The built FL algorithm suffers from slow learning convergence.  
			\\ \cline{2-9}&
			\cite{49}&	Edge data caching&	VFL&- &	Mobile devices&	Edge server&	A proactive content caching with FL in MEC. &	Communication cost for FL implementation has not been discussed. 
			\\ \cline{2-9}&
			\cite{50}&	Edge data caching&	CFL&DRL &	User equipments&	Edge/cloud server&	An edge caching scheme with DRL and FL. &	Scalability of the proposed FL scheme has not been verified.
			\\ \cline{2-9}&
			\cite{52}&	Edge content caching&	DFL&Transfer learning &	Mobile users&	BSs&	An edge content caching framework in UAV-based 6G networks. &	Security and privacy performances have not been evaluated. 
			\\ \cline{2-9}&
			\cite{53}&	Edge content caching&	DFL&- &	IoT devices&	Edge servers&	An edge content caching framework for IoT networks with blockchain. &	Blockchain cost from consensus running in FL has not been considered.
			\\ \cline{2-9}& 
			\cite{54}&	Edge content caching&	CFL&- &	Mobile users&	BS&	An edge content popularity prediction model with FL. &	The proposed model is not scalable with more edge servers/BSs.
			\\ \cline{2-9}
			\hline

	\end{tabular}}
\end{table*}

\begin{table*}
	\centering
	\caption{Taxonomy of FL-IoT services (continued). }
	{\color{black}
		\label{Table:FL_Services_next}
		\begin{tabular}{|P{0.5cm}|P{0.5cm}|P{1.5cm}|P{0.5cm}|P{01.1cm}|P{1.1cm}|P{1.3cm}|P{4cm}|P{4cm}|}
			\hline
			\textbf{Issue}& 	
			\textbf{Ref.} &	
			\textbf{Use case}&	
			\textbf{FL type}& 
			\textcolor{black}{\textbf{ML type}}& 
			\textbf{FL clients}& 	
			\textbf{FL aggregator}&
			\textbf{Key contributions}&
			\textbf{Limitations}
			\\
			\hline
			\parbox[t]{1.5cm}{\multirow{16}{*}{\rotatebox[origin=c]{90}{FL for IoT Attack Detection}}} & 
			\cite{57} &	Attack defense&	VFL&DNN &	IoT devices&	Cloud server&	A federated attack defense scheme in industrial IoT networks.&	Communication latency during the FL process with cloud has not been considered. 
			\\ \cline{2-9}&
			\cite{64}&	Attack detection&	HFL&Adversarial ML &	IoT devices&	Security gateway&	An FL-based approach for attack detection in IoT. &	Data privacy has not been investigated.
			\\ \cline{2-9}&
			\cite{66}&	Attack detection&	VFL&NN &	IoT gateways&	Data center&	An FL-enabled scheme for federated attack detection in Industry 4.0. &	Scalability in contexts of Industry 4.0 for the built FL scheme has not been considered. 
			\\ \cline{2-9}&
			\cite{67}&	Intrusion detection &	CFL&DNN &	Routers &	Data center&	An intrusion detection mechanism with FL for IoT networks. &	The considered FL model is still simple. 
			\\ \cline{2-9}&
			\cite{68}&	Intrusion detection &	CFL&- &	IoT devices&	Cloud server&	An FL model for scalable intrusion detection in IoT. &	The comparison between ML and DL-based approaches in FL process is missing. 
			\\ \cline{2-9}&
			\cite{69}&	Malware detection&	CFL&Random forest &	Android devices&	Data center&	An FL-based solution for malware detection in Android apps. &	The performance of learning convergence has not been presented. 
			\\ \cline{2-9}
			\hline
			
			\parbox[t]{1.5cm}{\multirow{10}{*}{\rotatebox[origin=c]{90}{FL for IoT Localization}}} & 
			\cite{71}&	Indoor localization&	CFL&NN &	Mobile users&	Data server&	An FL-based localization scheme for building scenarios. &	Attack analysis has not been provided. 
			\\ \cline{2-9}&
			\cite{73}&	Indoor localization&	CFL&- &	Mobile users&	Data server&	An FL-based localization scheme for mobile crowdsourcing. &	The feasibility of the proposed model has not been investigated on real localization systems. 
			\\ \cline{2-9}&
			\cite{74}&	Indoor localization&	CFL&DNN &	WiFI devices&	Data server&	A scheme for indoor localization with FL in WiFi networks. &	Performances on FL convergence and learning latency have not been provided. 
			\\ \cline{2-9}&
			\cite{75}&	Indoor localization&	CFL&DNN &	Mobile devices&	Data server&	A federated localization (FedLoc) model for mobile localization. &	Security for mobile FL-based localization has not been considered. 
			\\ \cline{2-9}
			\hline
			
			
			\parbox[t]{1.5cm}{\multirow{13}{*}{\rotatebox[origin=c]{90}{FL for IoT Mobile Crowdsensing}}} & 
			\cite{77}&	Mobile crowdsensing&	HFL&Regression tree &	Mobile devices&	Data server&	A secure FL model for mobile crowdsensing. &	Scalability of the proposed scheme has not been investigated. 
			\\ \cline{2-9}&
			\cite{79}&	Mobile crowdsensing&	HFL&Regression tree &	Mobile devices&	Cloud server &	An FL-based architecture for mobile crowdsensing. &	Communication with cloud for FL training is high.  
			\\ \cline{2-9}&
			\cite{80}&	Mobile crowdsensing&	VFL&- &	Mobile users&	Edge server &	An FL-based scheme for federated crowdsensing with incentive. &	Scalability and latency performances should be given. 
			\\ \cline{2-9}& 
			\cite{81}&	Mobile crowdsensing&	DFL&- &	IoT sensors&	 MEC servers&	An FL-based crowdsensing with incentive in MEC. &	Only utility simulations are given.   
			\\ \cline{2-9}&
			\cite{82}&	Mobile sensing&	DFL &- &	IoT devices&	 Cloud servers&	A distributed FL solution for mobile sensing in IoT. &	The given model is simple and hard to apply for other scenarios. 
			\\ \cline{2-9}&
			\cite{84}&	Mobile crowdsensing&	DFL&RL &	UAVs&	Task publishers&	A blockchain-based FL scheme for mobile sensing in UAV networks. &	Impacts of blockchain consensus have not been evaluated.
			
			\\ \cline{2-9}
			\hline
			
			
			\parbox[t]{1.5cm}{\multirow{13}{*}{\rotatebox[origin=c]{90}{FL for IoT Privacy and Security}}} & 
			\cite{408}&	IoT privacy preservation &	VFL&- &	Vehicles&	Data centre&	An FL-based scheme for privacy preservation in vehicular IoT. &	Convergence performance has not been validated. 
			\\ \cline{2-9}&
			\cite{409}&	IoT privacy preservation&	HFL&DNN &	Vehicles&	Cloud server&	An FL-differential privacy-based scheme for privacy enhancement in vehicular IoT. &	Communication latency during the FL process with cloud has not been considered.
			\\ \cline{2-9}&
			\cite{410}&	IoT privacy preservation&	DFL&NN &	Mobile users&	Companies&	A blockchained FL scheme for crowdsourcing systems. &	The impact of blockchain mining on FL training has not been verified. 
			\\ \cline{2-9}&
			\cite{411}&	IoT anomaly detection&	HFL&DNN &	Mobile users&	Edge server&	A multi-task FL scheme for IoT anomaly detection. &	The implementation is simple without detailed experiments. 
			\\ \cline{2-9}&
			\cite{413}&	IoT security protection&	DFL&CNN &	Mobile users&	Mobile users&	A blockchain-FL scheme for security enhancement in fog IoT. &	The joint learning and mining energy allocation should be considered. 
			\\ \cline{2-9}
			\hline
			
	\end{tabular}}
\end{table*}

%% file: FL_Applications.tex
\section{FL for  IoT Applications}
\label{Sec:FL_Applications}
\begin{figure*}
	\centering
	\includegraphics[width=0.97\linewidth]{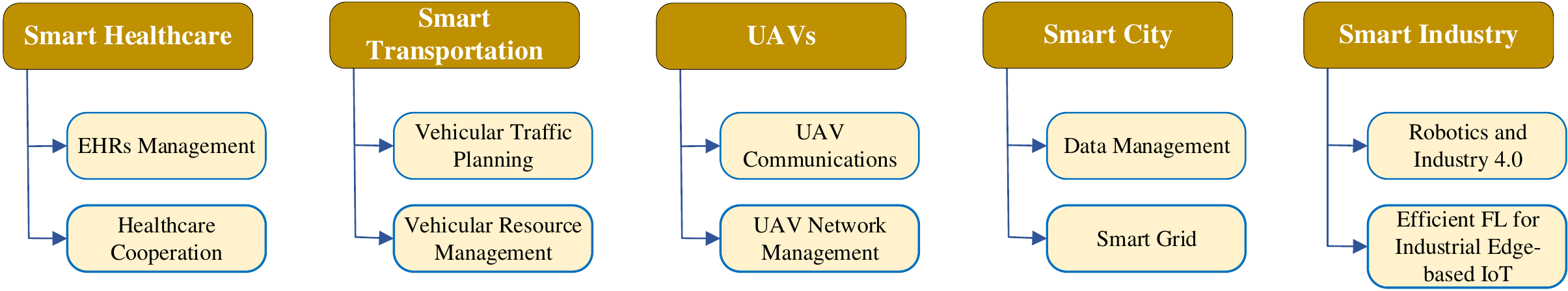}
	\caption{FL-IoT application domains. }
	\label{Fig:FL_ApplicationUseCase_Structure}
	\vspace{-0.1in}
\end{figure*}

{Enabled by the roles and benefits of FL in IoT services as presented in the previous section,} here we provide an extensive discussion of the integration of FL into a wide range of key IoT applications, including smart healthcare, smart transportation, UAVs, smart city, and smart industry along with applied use case domains, as summarized in Fig.~\ref{Fig:FL_ApplicationUseCase_Structure}.
\subsection{FL for Smart Healthcare }
In smart healthcare, AI-based approaches have been extensively adopted to learn health data to facilitate healthcare services, e.g., intelligent imaging for disease detection \cite{85}.  A key issue in such a traditional AI model is the privacy concern caused by the public data sharing with the cloud or the data centre for data training \cite{86}. Indeed, compared to other domains, data in healthcare systems are highly sensitive subject to health regulations such as United States Health Insurance Portability and Accountability Act (HIPPA) \cite{90}. The removal of metadata such as patient information is insufficient to preserve privacy of patients, especially in the complex healthcare settings where multiple parties such as hospitals and insurance companies have access to healthcare database {as a part of the employment requirements}, including data analysis and processing. Obviously, the use of traditional AI methods with the reliance on a central server for analytics is not an ideal solution for modern healthcare. FL can offer alternative solutions by providing intelligence with privacy awareness, where the data sharing is not needed \cite{87}. Recent works have demonstrated the practicality of FL in smart healthcare sector with advanced features. Here, we focus on analyzing the roles of FL in healthcare with two use cases, including EHRs management and healthcare cooperation. 
\subsubsection{FL for Electronic Health Records (EHRs) Management} 

Some research efforts have been devoted to the use of FL for enabling flexible and privacy-preserving EHRs management in healthcare operations. For instance, a collaborative learning protocol based on FL is introduced in \cite{88} for an EHRs system with the cooperation of multiple hospital institutions and a cloud server. Here, each hospital runs a NN using its own EHRs with the help of cloud server. To ensure privacy for model parameters in the FL process, a lightweight data perturbation method is considered to perturb the training-related data, which thus can {defend model memorization attack} in the learning. Although attackers can obtain perturbed information of EHRs, it is hard to obtain or recover the original data. Simulations on AlexNet NN with standard CIFAR-10 dataset verify a good performance in terms of prediction accuracy and security levels for EHRs learning. The authors in \cite{89} also build a federated NN training framework that allows each hospital participates in learning part of the model using its EHRs data source. An eICU collaborative research database collected from 58 hospitals comprising 1,264,89 ICU admissions is used to evaluate the proposed FL algorithm in terms of the prediction rate of patient mortality during the ICU admission. To save network resources spent by the communication to the central server in EHRs learning, the study in \cite{91} introduces a fully decentralized FL approach by combining classic non-convex decentralized optimization and decentralized stochastic gradient tracking. Each hospital runs a learning model locally to extract patients features from real-world datasets (i.e., proprietary clinical dataset including 7919 patients diagnosed with mild cognitive impairment) by using a decentralized stochastic gradient  algorithm combined with a linear speedup technique to accelerate the convergence rate. Although FL enables distributed learning without sharing EHRs, model updates still remain privacy leakage due to inference attacks on the communication channel. To overcome this challenge, differential privacy techniques can be useful to improve privacy protection for FL-based EHRs learning \cite{92}. By using an objective perturbation method which adds noise to the objective function at local models, we can acquire a differentially private approximation that produces a minimizer of the perturbed objective. Several simulations have been tested with using AI models such as gradient descent, namely perceptron, SVM, showing that the proposed approach can offer strong privacy levels while ensuring high training performance.

The potential of FL has been investigated in solving distributed binary supervised classification problems to estimate hospitalizations for cardiac events \cite{93}. Here, each data holder, such as a mobile user with smartphone, runs a SVM model using EHRs datasets featured by age, gender, and race and physical characteristics, and then exchanges the computed updates with an aggregator for building a global hospitalization prediction model. The proposed solution has the potential to support prediction of the progression of several popular diseases with cardiovascular conditions. FL is also applied to predict adverse drug reactions (ADR) on EHRs data to solve issues of data shortage in a single site for rare ADR detection \cite{94}. Each medical site performs an AI model such as SVM, single-layer perceptron, and logistic regression, and contributes to the computation of a global model by using its sensitive and imbalanced real-world EHRs data. Experiments for two use cases, namely prediction of chronic opioid usage and prediction of extrapyramidal symptoms for patients, are implemented. Compared to the centralized AI approaches, FL offers a similar accuracy performance in predicting ADR without sacrificing user data privacy. To further improve the accuracy rate and accelerate the convergence speed for FL in EHRs learning, the authors in \cite{95} suggest to remove irrelevant updates while exploring the relevance of local updates using a sign method at each EHRs owner in the FL architecture consisting of secret providers, EHRs owners, and a central server. Further, security and privacy are also considered in \cite{96} for an FL-based medical imaging processing architecture. Here, hospitals, healthcare providers and patients cooperatively train a secure FL model with the support of differential privacy to build a secure multi-party computation model for image analytics by an algorithm owner at a medical center.  
\subsubsection{FL for Healthcare Cooperation}

In addition, FL with its distributed and privacy-preserved nature can promote secure healthcare cooperation for better medical service delivery. The study in \cite{97} presents a cooperative healthcare framework enabled by FL among medical IoT devices. Each device joins to run a NN using electrocardiogram datasets to build an arrhythmia detection model at a powerful server. Compared to the FedAvg algorithm, the proposed FL scheme tested on 64 IoT devices can achieve a lower communication overhead with a small accuracy loss in a practical arrhythmia detection task. Meanwhile, in order to solve issues caused by device, data, and model heterogeneity that can adversely affect the FL process, a new personalized FL solution is proposed in \cite{99} for cloud-edge-based healthcare. In this case, personalization learning is performed at local devices to mitigate heterogeneities while attaining high-quality personalized models.
{Enabled by FL-based optimization for data offloading as presented in subsection~\ref{Subsection:Dataoffloading}, a federated offloading scheme is designed for mobile healthcare.} That is, each IoT device can choose to offload its computationally intensive tasks to edge gateways which execute the learning models before sending the updates to the cloud for combination. This realizes a cooperative network of cloud, edge, and IoT devices for healthcare applications with the help of differential privacy and homomorphic encryption techniques for security improvement. To promote the federation of wearable devices in supporting healthcare, a transfer learning framework called FedHealth is studied in \cite{100} for wearable healthcare with the help of FL. FedHealth allows to aggregate the data from separate hospital organizations with multiple wearable IoT devices to build a strong AI model for medical tasks, such as human activity recognition, enabled by homomorphic encryption for health data privacy preservation \cite{101}. By using the computational capability of distributed hospitals, FL offers a powerful data analytic solution for improving recognition accuracy rate compared to centralized AI approaches, as confirmed in numerical simulations. To reduce latency spent by communications among FL clients and the server, a new chain-directed Synchronous Stochastic Gradient Descent approach is introduced in \cite{102} for personal mobile sensing in healthcare applications. Based on a modified DL4J library integrated on smartphones, two CNNs are performed by employing multi-channel sensing data, indicating a high performance in terms of high training accuracy and communication delays reduced by 53\% with a Ring-scheduler approach. Smartphones have been utilized in \cite{103} to implement an FL algorithm for federated mobile healthcare, with a focus on solving the cold start issue caused by slow data generation and computation of certain mobile devices in the cooperative FL process. For large-scale healthcare cooperation, blockchain \cite{507dinh1} has been emerged as a viable solution that can be integrated with FL for building decentralized healthcare systems involving a large number of medical entities acted as data workers \cite{104}. By using blockchain, the central authority in the traditional FL architecture is eliminated, which promotes network connectivity and accelerates the training process over the large-scale healthcare system. Moreover, fine-grained data access policies on blockchain, such as smart contracts \cite{105}, can provide reliable authentication for federated health data processing. A similar decentralized approach on a P2P network for federated healthcare is also studied in \cite{106}. In this context, all data centers directly communicates with each other without relying on a central authority, which thus reduces data leakage due to curious third-parties and mitigates communication delays.  

\begin{figure}
	\centering
	\includegraphics[width=0.97\linewidth]{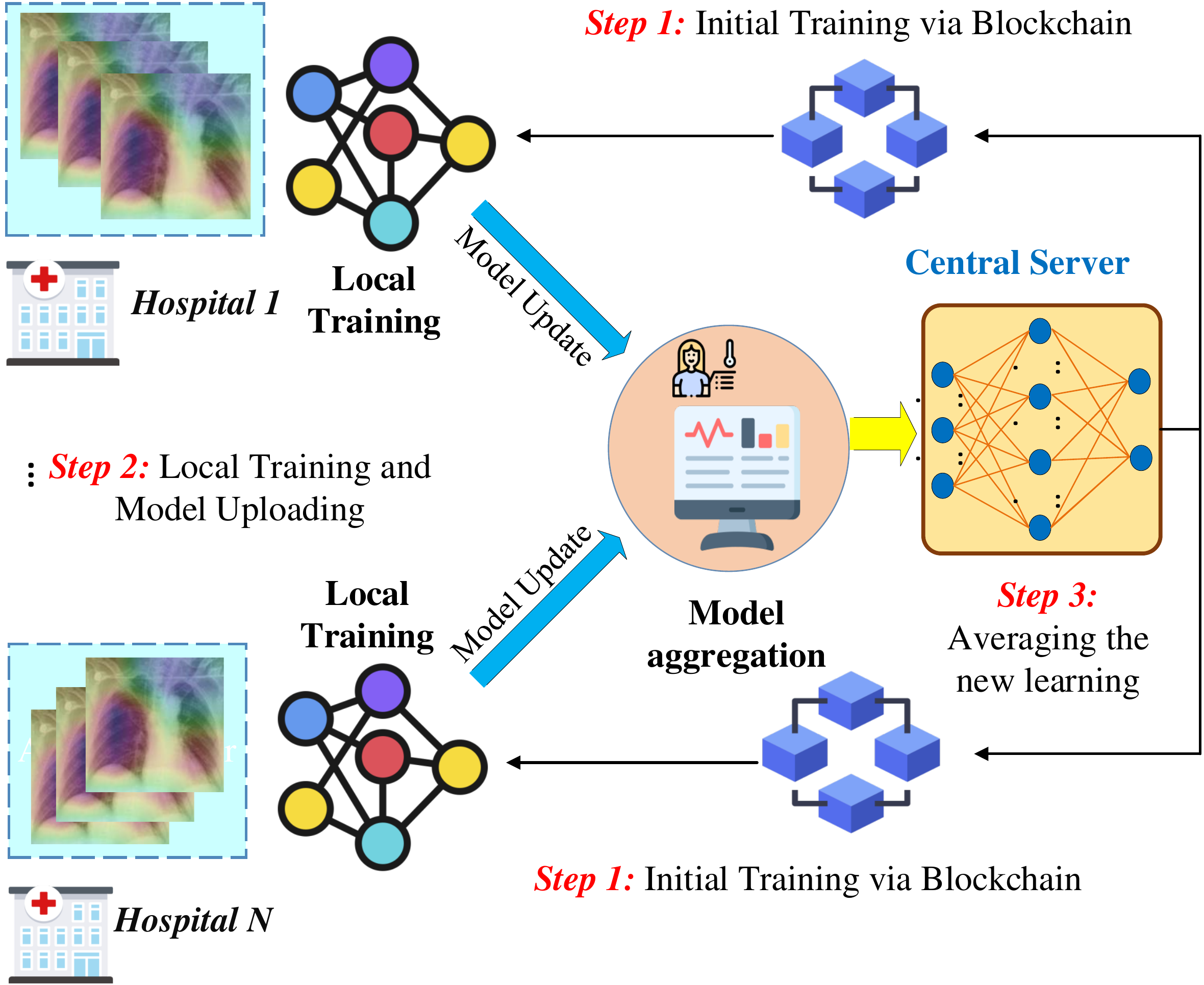}
	\caption{Federated learning for smart healthcare: A case study for COVID-19 image classification with blockchain \cite{108}. }
	\label{Fig:FL_Covid}
	\vspace{-0.1in}
\end{figure}

Very recently, the potential of FL to support the fighting against infectious diseases such as COVID-19 \cite{107} has been investigated \cite{108}, as indicated in Fig.~\ref{Fig:FL_Covid}. A number of hospitals cooperatively communicate each other via the blockchain to run DL algorithms locally for identifying CT scans of COVID-19 patients. At each hospital, a deep capsule network is developed to enhance image classification performance, while FL provides guidance for transmissions of model updates to perform the model aggregation at a common hospital. Simulations from 34,006 CT scan slices (images) of 89 subjects verify a high COVID-19 image classification and low data loss in FL algorithm running.  FL is also used in \cite{109} to provide privacy-preserved AI solutions for COVID-19 chest X-ray image analytics. Several practical experiments have been implemented where multiple COVID-19 CXR image owners run local learning networks such as ResNet18 for image classification and then share the computed parameters with a data center for mobile averaging while the data ownership of each user is ensured. 
\subsection{FL for Smart Transportation}
In the past few years, AI/ML techniques have been exploited to enable intelligent transportation systems (ITSs) by performing centralized vehicular data learning at the data centre \cite{110}, which requires data sharing in untrusted environments and thus potentially introduces privacy issues. FL has been introduced to bring AI functions to the network edge, involving a number of participants, such as vehicles, to collaboratively train globally shared AI models without the need for long data transmission and compromising user privacy. Reviewing the literature in the FL-based smart transportation domain, we here focus on analyzing the applications of FL in vehicular traffic planning and vehicular resource management.  
\subsubsection{FL for Vehicular Traffic Planning}

Many FL-based architectures have been proposed to support vehicular traffic planning which is an important service in ITS for traffic prediction and vehicle control for congestion minimization. For example, FL is considered in \cite{9} to replace traditional centralized ML approaches in traffic prediction tasks by running ML models directly at the edge devices, e.g., vehicles, based on their datasets such as road geometry, traffic flow and weather. Another privacy-aware traffic prediction solution is studied in \cite{112} where multiple entities such as government, companies, stations join to run a Gated Recurrent Unit neural network (FedGRU) locally to estimate traffic flow and then calculate local updates for aggregation at a data center. Here, an improved FedAvg algorithm is developed based on a joint announcement-enabled an aggregation mechanism for better scalability of the FL scheme. Simulation results from using a Caltrans Performance Measurement System (PeMS) database \cite{113} show less accuracy degradation and high privacy levels over centralized learning methods. In \cite{114}, a traffic simulator is introduced with the help of FL for guiding reinforcement learning (RL) agents on vehicles in self-driving tasks by pooling their robotic car resources without sharing raw data. The proposed scheme is promising to support collision avoidance RL tasks in high-speed autonomous driving where latency and privacy are among the critical concerns. To attract more vehicles to run computation algorithms for traffic prediction, an incentive mechanism is designed in \cite{115} for UAVs-based vehicular networks. In this scenario, UAVs are used to provide data collection and computation offloading support for Internet of Vehicles (IoV), such as capturing car parking and monitoring traffic conditions from stationary vehicles and roadside units, and then participate in the privacy-preserved collaborative model training. A contract is designed among six UAVs and a single subregion to ensure the highest utility and the lowest marginal cost of each UAV within coverage areas. Meanwhile, the authors in \cite{116} combine FL with blockchain to build decentralized traffic planning solutions for vehicular systems. Each vehicle acts as an FL client to run an ML model and exchange computed updates together via a blockchain ledger while verifying their corresponding rewards. The use of blockchain potentially overcomes the challenges faced by traditional FL approaches in terms of long communication and security risks due to curious third-parties.  
\subsubsection{FL for Vehicular Resource Management}

In addition to traffic planning, FL has the potential to facilitate resource management strategies for vehicle-to-vehicle (V2V) networks. An FL-based approach is introduced in \cite{118} to support vehicular ultra-reliable low-latency communication (URLLC) where each vehicle can learn a generalized Pareto distribution (GPD) of network queues with respect to power control and resource allocation without revealing the information of queue length samples. Then, the computed GPD parameters are updated to the RSU for aggregation in a longer time scale. Compared to centralized learning approaches, FL offers more benefits for intelligent vehicular services such as better efficiency in resource usage, lower power consumption with a similar learning accuracy. Another resource allocation scheme for vehicle-to-everything (V2X) communication is also considered in \cite{119}. In this setting, FL is combined with DRL \cite{120} to build a federated intelligent resource allocation strategy, in order to maximize the sum capacity of vehicular users with respect to latency and reliability conditions. Each vehicular user is regarded as a DRL agent in the FL architecture to run a DNN algorithm for optimal mode selection and resource allocation, while the BS aggregates the updates offloaded from users to build undirected graphs using channel gain information. {Further, motivated by FL-based optimization for IoT data caching as discussed in subsection~\ref{Subsection:DataCaching}, a federated solution for caching and computing resource management in MEC-based vehicular networks is proposed in \cite{45}.} In particular, vehicles cooperatively communicate with RSUs to perform federated learning in which each entity computes a sub-gradient descent update as part of the joint parameter optimization for system cost minimization. Simulations with five RSUs and several vehicles demonstrate a high performance regarding learning accuracy compared to non-cooperative learning approaches \cite{123}. The authors in \cite{124} also build a federated Q-learning algorithm for cooperative tasks offloading with multiple V2X, aiming to optimize the failure probability and communication resource usage. A consensus Q-table is designed to guide the Q-learning agent to make decisions, e.g., offloading selection, in a federated manner for minimizing running cost over the vehicular network.  
\subsection{FL for Unmanned Aerial Vehicles (UAVs)}
UAVs play an important role in various services such as goods delivery, disaster monitoring or military and have been regarded as a key technology in 5G/6G wireless networks \cite{125} due to its high flexibility and seamless connectivity. To provide intelligence in UAV networks, AI/ML techniques have been used for UAV tasks at ground BSs such as trajectory planning, power control, target recognition \cite{126}. However, because of the high mobility and high altitude of UAVs, it is challenging to ensure the continuous communications between UAVs and BSs with respect to dynamic aerial environment conditions. Therefore, using centralized AI/ML approaches to perform UAV tasks may not be an ideal solution, especially when having to transmit a large amount of data over the aerial links. Distributed learning, such as FL, can provide better learning solutions for intelligent UAVs networks by using a cooperation of multiple UAVs without transferring raw data to BSs and sacrificing data privacy. We here analyze the use of FL in UAVs via two aspects, including UAV communications and UAV network management.  
\subsubsection{FL for UAV Communications}

Several works have been done on the use of FL for facilitating UAV communications. The study in 
\cite{127} leverages FL to build a federated path control strategy for large-scale UAV networks. Each UAV runs a NN and then share the model parameters with a central unit for obtaining a global model, which makes the estimation of the population density function at UAVs more accurate. The use of FL would mitigate the data volume transferred over the aerial environment and thus reduce communication delays and privacy concerns, while increasing training speed of the global model due to the employment of computing resources of all UAVs. The simulations confirm that federated learning algorithms can reduce transmission time, motion energy of UAV communications, as well as achieve minimum collision risks in windy environments. Instead of using a central server like in \cite{127}, the authors in \cite{128} use a leading UAV as an FL aggregator to manage a swarm of following UAVs in an intra-swarm network. Based on a defined minimum number of communication rounds, the key objective is to jointly optimize both power allocation and scheduling for UAVs, aiming to reduce the FL convergence round, with respect to learning, communication delay, and flying coverage constraints. Through a numerical simulation, the proposed joint scheme can reduce the convergence round by 35\%, compared to non-joint schemes (i.e., power allocation or scheduling design). Another work in \cite{129} pays attention to federated beamforming design \cite{130} for UAV communications, by employing a local extreme learning machine (ELM) model with respect to CSI consideration. Then, a stochastic parallel random walk alternating direction algorithm is designed based on UAV dynamics and CSI collection to accelerate the convergence rate to a consensus among UAVs.  Another work in \cite{131} focuses on developing a convolutional auto-encoder based on FL for illumination distribution in UAV communications. Unlike the traditionally centralized ML techniques that perform all ML learning at a server with a complete set of illumination data, FL enables UAVs to collaboratively train the auto-encoder only by using their partial illumination data, which potentially reduces transmission power and improves privacy. In this way, UAVs have more flexible solutions to adjust its serving position and user association for communication power savings. Toward UAV-based 6G networks, FL has been used in \cite{132} for on-demand 3D deployment by connecting UAVs with a BS.  In particular, an air-to-air FL algorithm is built that allows UAVs to train their model within the aerial environments, with the help of controllable deployments of cooperative UAVs for low communication energy consumption while remaining high learning accuracy. 
\begin{figure}
	\centering
	\includegraphics[width=0.97\linewidth]{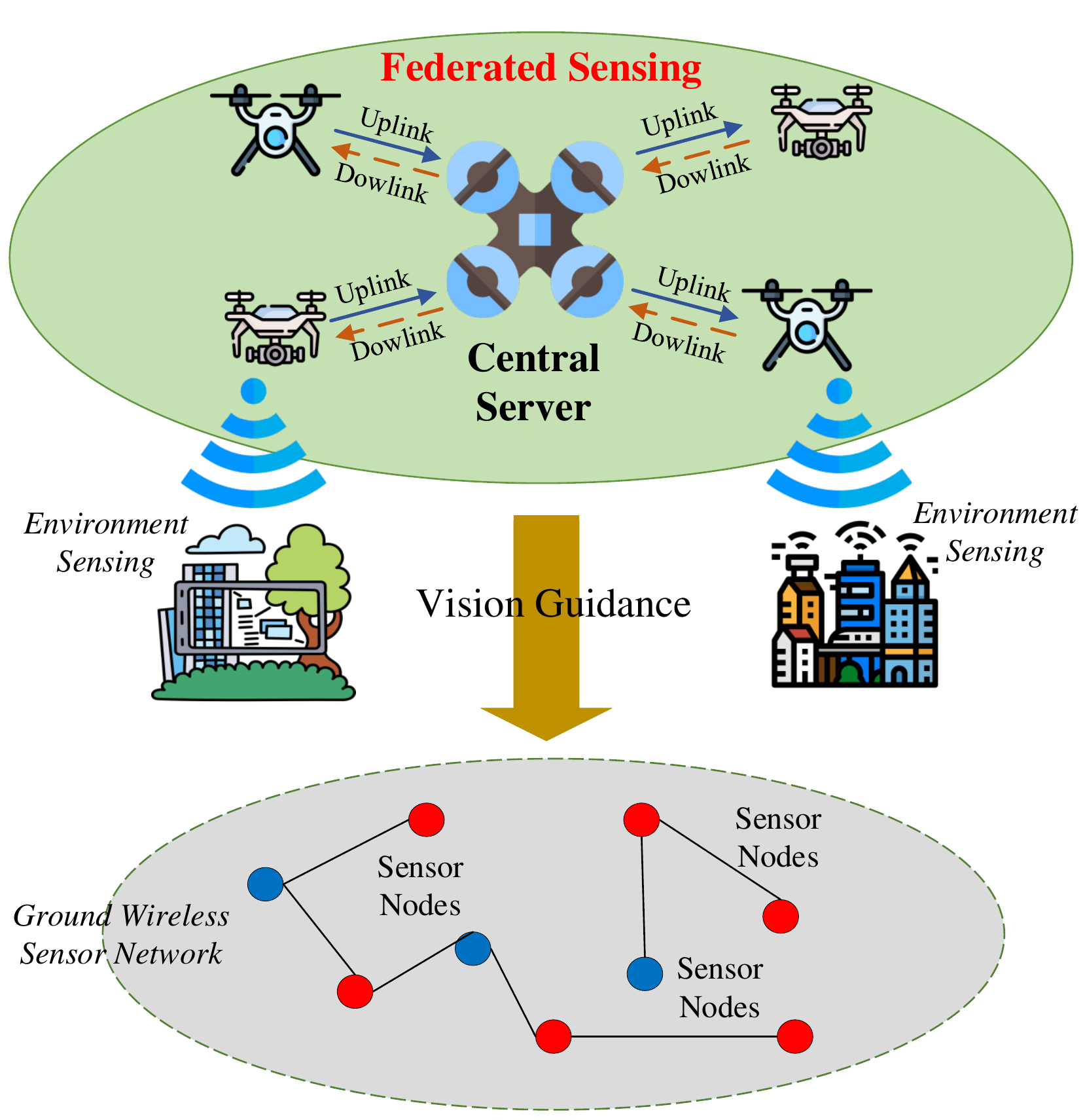}
	\caption{Federated learning for UAV-based ground air quality monitoring \cite{133}. }
	\label{Fig:FL_UAV_Network}
	\vspace{-0.1in}
\end{figure} 
\subsubsection{FL for UAV Network Management}
The work in \cite{133} studies a federated UAV management architecture for aerial UAV swarms sensing network integrated with a ground sensing network for forming a hybrid spatio-temporal sensing system, as shown in Fig.~\ref{Fig:FL_UAV_Network}. Each UAV acts as an FL client to monitor the air quality index in its coverage area without revealing raw data by cooperating UAVs with a central server which is responsible for running a light-weight DenseMobileNet model based on haze features offloaded from distributed UAVs. The ultimate goal is to provide air quality forecasts while controlling UAV energy consumption in aerial sensing tasks. Compared to CNN and SVM-based algorithms, the proposed federated learning scheme can achieve a better accuracy in air quality estimation with privacy protection and energy efficiency. {Based on the FL concept for IoT attack detection in subsection~\ref{Subsection:AttackDetection}, a federated UAV network management architecture is also considered in \cite{134} for solving issues related to security threats.} To be clear, a federated jamming attack detection approach is introduced that allows UAV clients to perform AI model training locally with respect to communication efficiency and unbalanced data properties before sharing computed updates to a server based on a Dempster-Shafer theory-based client group prioritization model. By using a CRAWDAD jamming attack dataset, the learning accuracy for jamming detection is improved with low training time in various UAV group evaluations. A similar security protection scheme is presented in \cite{136} where FL is combined with reinforcement learning for defense strategies against jamming attack. Learning updates from FL models at UAVs are combined by a Q-learning table using a Bellman equation \cite{137} that is able to determine optimal UAV paths for minimal security risks. The application of the adaptive federated reinforcement learning approach thus can achieve high accuracy rate in attack detection (over 82\%) with fast convergence rate and high learning rewards.  
\subsection{FL for Smart City }
The integration of smart devices and high-tech infrastructure, and integrated monitoring systems in city environments along with communication technologies forms smart city ecosystems, aiming to promote the quality of life for urban citizens, by enabling the seamless delivery of food, water, and energy to end users \cite{138}. To realize smart cities, AI/ML techniques have been widely adopted to provide intelligence properties thanks to their ability to handle real-time big data generated from sensors, devices, and human activities \cite{139}. To do that, most of AI-based smart city solutions rely on a centralized learning architecture on a data center, such as a cloud server, which is obviously not scalable to the rapid expansion of smart devices in smart cities. FL offer more attractive features for enabling decentralized smart city applications with high privacy levels and low communication delays. There are two key domains FL can offer useful services for smart city, including data management and smart grid.  

\subsubsection{FL for Data Management}

With its decentralized and privacy-preserved nature, FL has been exploited to provide distributed AI functions for large-scale intelligent data management systems in smart cities. For example, a semi-supervised FL method called FedSem is introduced in \cite{140} to provide distributed processing for unlabeled data in smart cities. To evaluate the usefulness of FL in a smart city, a prototype with smart vehicles is considered where each vehicle learns a DNN model based on traffic sign image datasets. Then, a central server coordinates to select multiple participants in each learning round for model orchestrating. A German traffic sign dataset between 1000 vehicles is employed to simulate the FL algorithm in a fashion in each round, only 30 vehicles are selected to train and perform model updating. Implementation results indicate a good performance in terms of high accuracy without much testing loss for unlabeled smart city data. To overcome the challenges posed by the ubiquity of smart city data from different streams and devices and user privacy concerns, FL is also used in \cite{142} to structure data streams from ubiquitous IoT devices that act as FL clients to perform local learning without sharing their data to external third-parties. This would reshape the current forms of smart cities by providing newly exiting services such as smart urban communication, social economy sharing, social activity monitoring, and interconnection of global citizens \cite{143}. These services can be empowered by intelligent sensing which allows IoT devices to sense physical environments and perform data learning for extracting useful information to serve end users \cite{82}.  In this regard, FL can be employed to develop distributed sensing platforms with pervasive computing for smart cities. Each device can collect data locally and runs part of an AI model such as a NN, using its hardware without exchanging its personal information for security and privacy. As a result, communication costs such as latency are significantly reduced while learning qualities are ensured. FL for mobile pervasive computing in smart cities is also considered in \cite{145}. For example, vehicles can join an FL system to collaboratively train AI models for prediction of locations of charger installations without revealing information to RSUs. This kind of on-device processing offered by FL thus can help solve issues related to data ethics, data privacy, and security in smart city services. Meanwhile, the authors in \cite{36} suggest to use FL for building a video data management platform in smart cities. To be specific, videos can be collected as live street videos from connected mobile cameras such as IoT devices on buses, roads and transferred to edge devices. Then, each edge device runs a semi-supervised learning algorithm that can perform local video analytics based on multiple video segments divided from raw video frames. To solve the issue of the non-IID data, a FedSwap operation solution is proposed, aiming to mitigate the diversity of the data and increase the accuracy of image classification by 3.8\%, as confirmed in simulations.  
\subsubsection{FL for Smart Grid}

Smart grid holds an significant role in supporting industrial systems and manufacturing processes, by delivering energy resources to the households and factories via electrical grids \cite{147}. FL can provide privacy-enhanced intelligent solutions to realize safe smart grid operations.  Indeed, FL is used to establish federated predictive power schemes in a network of edge data centers that run a recurrent neural network locally to estimate future energy demands based on historical customer energy usage datasets before being aggregated at a cloud server for constructing a global prediction model \cite{148}. In this way, user information such as energy preference and home addresses is not revealed to the cloud for privacy protection. Moreover, by cooperating data centers over different urban areas, the prediction model can achieve a high learning performance with better accuracy rate, compared to centralized learning solutions at a single server \cite{149}. Another FL algorithm is also designed in \cite{151} for electricity power learning in power IoT networks consisting of electric providers and IoT users. A communication model is then formulated based on an FL process that aims to solve the tradeoff between resource consumption, user utility and local differential privacy.  

\subsection{FL for Smart Industry}
\begin{figure*}
	\centering
	\includegraphics[width=0.9\linewidth]{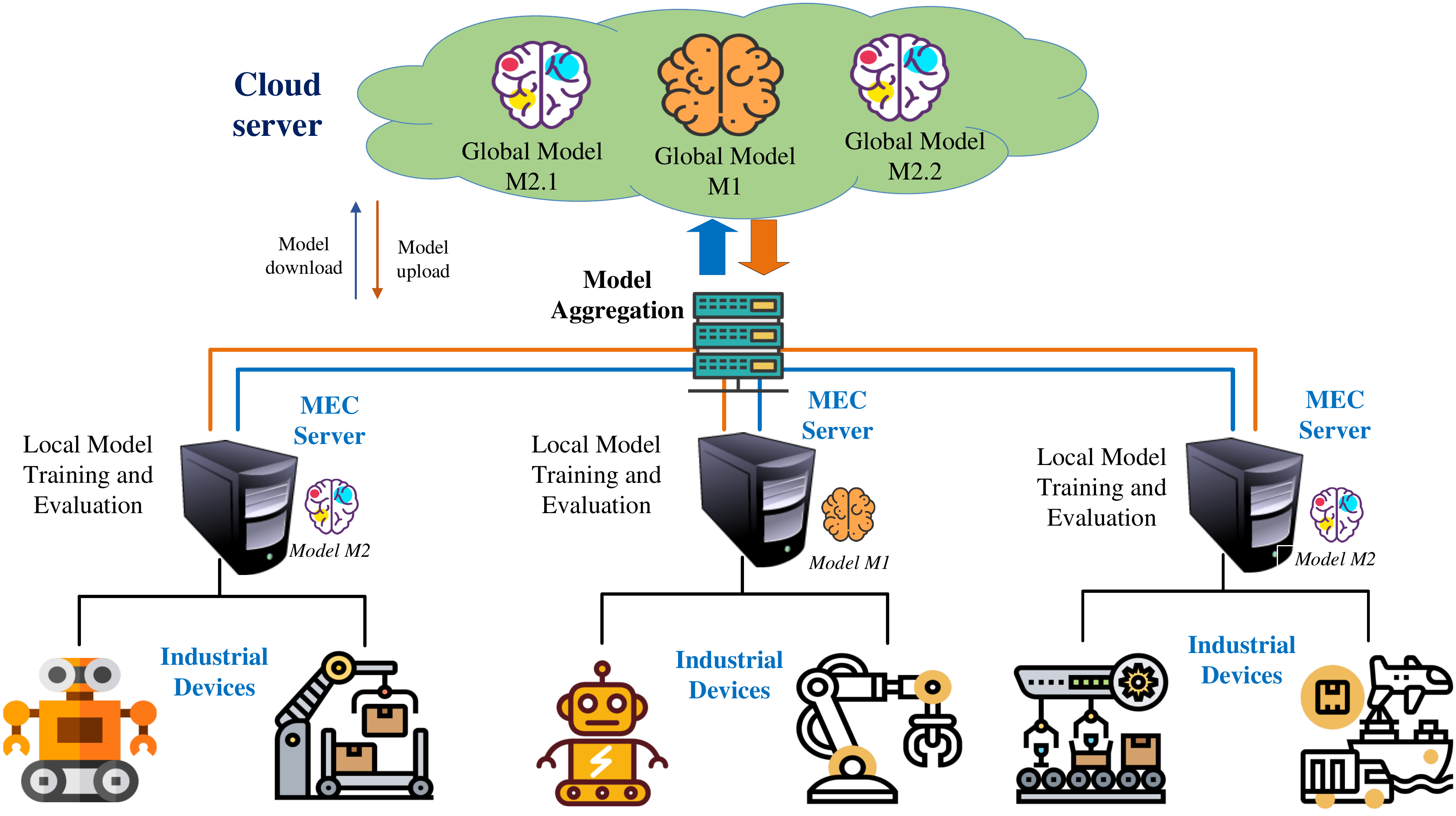}
	\caption{Federated learning for cooperative industry \cite{153}. }
	\label{Fig:FL_Industry}
	\vspace{-0.1in}
\end{figure*}

Smart industry refers to the integration of intelligence into manufacturing processes where AI techniques such as ML and DL play important roles in learning big data generated from industrial machines for process modeling, monitoring, prediction and control in production stages \cite{152}. The performance of AI functions mostly depends on available training data, but it requires sharing among companies and factories. Due to the raising concerns of user privacy issues, sharing a large amount of data over the industrial networks for AI implementation is not an efficient solution. FL has emerged as a much more attractive approach that helps realize intelligence for industrial systems without data exchange and privacy leakage \cite{153}, as shown in Fig.~\ref{Fig:FL_Industry}. Here we focus on analyzing the roles of FL in robotics and Industry 4.0, and then discuss efficient FL solutions for industrial edge-based IoT. {Then, we introduce several real-world FL implementation cases and testbeds in industrial IoT.}
\subsubsection{FL for Robotics and Industry 4.0}

Robotic is a crucial component of automobile industrial systems that is able to handle manufacturing tasks by its automated and programmable features.  How to perform real-time data processing and protect data privacy for robotic systems is a critical challenge. FL can help solve these problems by allocating intelligence to robotic devices, instead of relying on a remote server for data processing. For example, FL can be adopted to support robotics in data learning by implementing AI models at local robots without unpredictable network transmission delays \cite{154}. In this way, each robot only offloads the gradient parameters updates to build a shared model at the cloud without sharing its raw data for privacy guarantees based on a differential privacy technique. The authors in \cite{155}, \cite{156} investigate a federated imitation learning scheme for cloud robotics. Each robot participates in training an imitation NN using its own sensor image dataset. Then, it offloads the updated parameters to the cloud for fusing knowledge before sending back to robots for the next round of learning and thus, robots can benefit from exchanged knowledge values. Through communication rounds, the server can accumulate a significant knowledge of different robots to build a powerful learning model, and thus the imitation learning efficiency and accuracy of local robots can be improved in comparison with centralized learning approaches. As an extended version of cloud computing, fog/edge computing is also helpful to provide low-latency communication for federated robotics \cite{157}. Both computing, networking, and storage resources among robots are shared with a fog server for enabling federated learning with security awareness. Edge computing is also integrated with FL in \cite{158} to realize a collaborative learning among robotic arm devices. Due to the complexity and dynamics, each device runs a separate reinforcement learning model for determining its own control policy, and then shares mature policy model parameters to a cloud server for combination. Experiments on multiple rotary inverted pendulum devices \cite{158} confirm high learning performance regarding learning speed with respect to different learning clients. {Enabled by the FL-based IoT localization concept as presented in subsection~\ref{Subsection:IoTLocalization}, a federated Simultaneous Localization and Mapping (SLAM) system with cloud computing is deployed in \cite{159}.} Three robots coordinate with a cloud server to build an FL algorithm, aiming to construct a global map of an unknown environment enabled by a deep learning detector that can achieve robust feature extraction and high feature matching accuracy. 

In the industrial revolution, Industry 4.0 is the new concept whose aim is to promote the automation of manufacturing and industrial operations, which stimulates the development of smart factories for producing smart products without human involvement \cite{160}. FL can be used to support distributed intelligent industry 4.0 systems. As an example, a privacy-preserved FL framework is introduced in \cite{161} to realize industrial intelligence where multiple mobile users join to build a common AI model based on local gradients that are encrypted with homomorphic ciphertext with a distributed Gaussian mechanism at a cloud server. This solution potentially reduces the risks of privacy leakage from the local gradients and shared parameters in the FL communication rounds. To enhance the performance of compound TCP in industry 4.0 WiFi networks, the authors in \cite{162} suggest to use FL to collaborate multiple access points with respect to WiFi uploading and downloading dynamics and wireless data losses. Each access point performs a regression model based on quality of service (QoS) data samples and specifies learning parameters for contributing to a global server. Recently, an integrated FL-blockchain architecture is designed for industrial IoT networks \cite{163}. FL provides the ability of distributed learning at local IoT devices by using both static data and data streams, while blockchain can provide high degrees of security with immutable ledgers and smart contracts that can provide authentication for learning interactions in FL implementation. Blockchain is also employed in \cite{164} to provide security for FL implementation in industrial 4.0 networks. By using linked transactions, learning updates are verified and transmitted effectively for model aggregation, which in return accelerates learning convergence for FL algorithms.  
\subsubsection{\textcolor{black}{Efficient FL for Industrial Edge-based IoT Networks}}
\textcolor{black}{In recent years, industrial edge computing has been applied widely in IoT networks due to its capability to provide instant computation and storage services \cite{400}. The introduction of edge computing in IIoT can also significantly reduce the decision-making latency and save bandwidth resource. To enable intelligent and privacy-enhanced edge-based IIoT applications, efficient FL solutions have been applied by using the distributed learning capability of edge devices and the training federation in IIoT networks from both communication and network resource aspects.}

\textcolor{black}{- \textit{Communication-efficient FL for industrial edge-based IoT:} The authors in \cite{401} consider a communication-efficient FL approach called CE-FedAvg which is able to reduce the number of rounds to convergence and the total data uploaded per round over the traditional FedAvg scheme. This is realized by using a joint design of distributed Adam optimization and compression of uploaded models. In each communication round, the model server selects a subset of clients based on their power/communication properties and sends the weights and moments to them. Then, the server dequantizes and reconstructs the sparse updates from the clients and averages the updates for a global model that is downloaded by clients for sparsifying and quantizing the model deltas and sparse indices for compression. Implementation results on an industrial edge computing-like testbed using Raspberry Pi clients show the effectiveness of the proposed scheme with lower communication latency while the training accuracy is preserved. To solve the communication overhead issues for parameter synchronization in FL, a general gradient sparsification (GGS) framework is proposed in \cite{402} for FL-based edge IoT networks. The key idea is to combine gradient correction and batch normalization which allows the FL optimizer to properly treat the accumulated insignificant gradients, which makes the model converge better. It also mitigates the impact of delayed gradients on the global training without increasing the communication overhead. \textcolor{black}{Also, a dynamic FL solution is proposed in \cite{600} for power grid-based MEC environments where industrial IoT nodes such as metering devices collaborate with an MEC server to build an ML model for. To solve the issue of failure of communications between industrial clients and the server, a delay deadline constrained-FL framework is considered to avoid extremely long training delay via a dynamic client selection problem formulation for maximizing the computing utility with communication latency reduction in the FL process.} Another solution named communication-mitigated federated learning (CMFL) is introduced in \cite{403} for reducing communication overhead for FL-based edge IoT networks.  CMFL provides clients with the feedback information regarding the global tendency of model updates, aiming to identify the relevance of an update at a client with others. To do so, in each learning iteration, a client first receives the feedback information about the global update from the central server. Next, the client proceeds its local training and produces a local update. The client then compares it with the global update, checking how well the two gradients align with each other. By avoiding uploading those irrelevant updates to the server, CMFL can substantially reduce the communication overhead while still guaranteeing learning convergence.}

\textcolor{black}{- \textit{Network resource-efficient FL for industrial edge-based IoT:} Network resource optimization and allocation are important  to ensure robust FL training over distributed edge IoT networks. For example, a fair network resource allocation framework in wireless networks is suggested in \cite{404} that encourages a more  uniform resource (e.g., bandwidth) distribution across devices in FL-based edge networks. This allows for minimizing the aggregated reweighted loss  such that the devices with higher loss are given higher relative weight, by taking into account the important characteristics of the federated setting such as communication-efficiency and low participation of devices. The work in \cite{405} focuses on formulating a joint computation and transmission optimization problem, aiming to minimize the total energy consumption for local computation and wireless transmission in FL-based edge IoT systems. To solve this problem, an iterative algorithm is proposed with low complexity where in each step of this algorithm, new closed-form solutions are derived for the joint optimization of time allocation, bandwidth allocation, power control, computation frequency, and learning accuracy. In FL-based edge networks, how the  model  owner  can decide amounts of energy recharged to the workers and to choose channels, i.e., the default channel or the special channels, for  global  model  transmissions  to  maximize  the  number  of successful transmissions while minimizing the energy cost is highly important. To overcome these challenges, the work in \cite{406}  proposes   to   employ a DRL algorithm that  enables  the  model  owner  to  find the optimal decisions on the energy and the channels with no existing  network  knowledge.  A  stochastic  optimization  problem  of  the  model  owner is derived that maximizes  the  number  of  global  model  transmissions  while minimizing  the  energy  cost  and  the  channel  cost. \textcolor{black}{Furthermore, an industrial IoT-based FL scheme is also studied in \cite{601} with a focus on resource management for mobile IoT devices such as robots by considering multiple resource constraints, e.g., bandwidth, processor, or battery life which has not been considered in previous FL schemes with assumptions of sufficient resource availability. In each communication round, the resource availability is checked for model training and the trust score of each robot is verified to make decision for client training. Regarding the robots with delayed model update response,  the trust score can be adjusted in the next round of training, which potentially reduces the straggler effect in robotic-based FL systems.}}

\subsubsection{{FL Implementation and Testbeds in industrial IoT}}
{Inspired by the great potential of FL in IoT systems, there are several recent projects implemented to investigate the feasibility of FL in real-life industrial applications. As an example, the work in \cite{testbed1} implements a testbed for an FL-based smart home platform in a real-world IoT setting. More specifically, a smart home architecture is proposed, consisting of smart home IoT devices (e.g., camera, light bulb, door locks), a router, and an intrusion detection system with a SQLite database. By using an FL algorithm, smart home devices can train an ML model using their local information to share the learned models with the router for combination. In this way, FL helps users (i.e., smart home owners) to build home assistant solutions such as object detection and control their home by the coordination of distributed IoT devices in a privacy-preserving manner. Another project in \cite{testbed2} focuses on a verifiable FL platform to achieve efficient and secure model training in industrial IoT. A verification mechanism is built at the FL clients (i.e., industrial IoT devices) to verify {the accuracy of the aggregated results} based on the characteristics of Lagrange interpolation, which allows devices to detect forged results in the FL training. The proposed FL model is promising to solve secure training issues in intelligent industrial applications. For example, FL can be used to train risk assessment models in enterprise risk assessment tasks, by collaborating multiple banks to build high-quality risk assessment models without leaking the enterprise customer data. FL is also tested in \cite{testbed3} to provide privacy-preserved intelligence for cyber physical systems smart farming and smart logistics in industries. A platform called FengHuoLun is designed involving three views from macro to micro, namely entity view, edge view, and global view. Here, the entity view includes industrial IoT devices; the edge view implements the business requirements from industrial stakeholders, while the global view is at the cloud layer for running an FL algorithm that aggregates distributed ML models computed at local devices in the entity layer. A testbed is implemented in a wireless sensor network for intelligent abnormal detection with sparse representation, but experimental results have not been reported. In \cite{46}, an FL-based system is integrated in fog environments where collaborative IoT devices (e.g., smart factory machines) perform local data processing and transmit their learned parameters to a cloud server for model aggregation after each time interval. A testbed is implemented on a MNIST dataset combined with data traces from a Raspberry Pi to simulate the network delays and resource usage when integrating FL in IoT networks. The implementation results verify the efficiency of FL in terms of low communication latency due to model learning without sharing raw data and privacy preservation with guaranteed training accuracy. Several other works also investigate the practicality of FL in real-world healthcare settings. For instance, a federated edge learning system called FEEL is designed in \cite{testbed5} for mobile healthcare. Here, an edge-based training task offloading strategy is proposed to improve the training efficiency at distributed health users, while a differential privacy scheme is integrated to strengthen the privacy preservation during the FL training. A real-world healthcare FL experiment is tested in a network of 100 hospitals with using clump thickness as physiological attributes for building training samples, showing a low resource consumption and good privacy protection. }


In summary, we list FL-IoT applications in the taxonomy Table~\ref{Table:FL_Applications} and Table~\ref{Table:FL_Applications_next} to summarize the technical aspects as well as the key contributions and limitations of each reference work.
\begin{table*}
	\centering
	\caption{Taxonomy of FL-IoT applications.}
	{\color{black}
		\label{Table:FL_Applications}
		\begin{tabular}{|P{0.5cm}|P{0.5cm}|P{1.5cm}|P{0.5cm}|P{0.5cm}|P{1.2cm}|P{1.5cm}|P{4cm}|P{4cm}|}
			\hline
			\textbf{Issue}& 	
			\textbf{Ref.} &	
			\textbf{Use case}&	
			\textbf{FL type}& 
			\textcolor{black}{\textbf{ML type}}& 
			\textbf{FL clients}& 	
			\textbf{FL aggregator}&
			\textbf{Key contributions}&
			\textbf{Limitations}
			\\
			\hline
			\parbox[t]{1.5cm}{\multirow{26}{*}{\rotatebox[origin=c]{90}{FL for Smart healthcare}}} & 
			\cite{88} & 	EHRs management &	HFL & DNN &	Hospitals &	Cloud server &	A collaborative learning with FL for EHRs processing. &	Convergence of the FL algorithm has 	not 	been verified.  
			\\ \cline{2-9}&
			\cite{89} & 	EHRs management&  	HFL & NN &	Hospitals &	Data server&  	An FL scheme for EHRs learning in 
			distributed hospitals. &	The proposed model is simple and lacks 
			detailed evaluations.  
			\\ \cline{2-9}&
			\cite{91} & 	EHRs management &	DFL & NN &	Hospitals &	Hospitals &	A 	fully decentralized 
			EHRs learning with FL. &	Privacy of EHRs has 	not 	been considered.  
			\\ \cline{2-9}&
			\cite{93} & 	EHRs management &	DFL & SVM &	Health users &	Data center &		An 	FL-based 
			distributed binary supervised classification scheme. &	Simulation results for FL implementation have not been reported.  
			\\ \cline{2-9}&
			\cite{94} &	EHRs management&  	DFL & SVM &	Medical 
			sites &	Data center&  	An FL-based scheme for drug reaction 
			prediction 	on 
			distributed EHRs. &  	The 	healthcare scalability of the proposed method has not been considered.  
			\\ \cline{2-9}&
			\cite{95} &	EHRs management &	VFL & - &	EHRs owners &	Data center&  	A secure FLempowered 
			scheme 	for EHRs 
			computation. &	Accuracy of the FL 	algorithm has 	not 	been verified.  
			\\ \cline{2-9}&
			\cite{97} &	Healthcare cooperation &	HFL & NN &	Healthcare IoT devices &	Data server&  	An efficient FLbased model for healthcare. &	IoT user privacy has not ben taken into consideration. 
			\\ \cline{2-9}&
			\cite{99} &	Healthcare cooperation &	HFL & CNN &	IoT devices&  	Cloud server &	A 	cloud-edgebased healthcare model with FL. &	Attack analysis for FL has not been provided.   
			\\ \cline{2-9}&
			\cite{100} &	Healthcare cooperation&  	FTL & CNN &	Hospitals  &	Cloud server &	A scheme for 
			federated activity recognition with FL. &	The 	scalability with 	more hospitals has not been investigated.  
			\\ \cline{2-9}&
			\cite{102}&  	Healthcare cooperation&  	CFL & RL &	Smartphones&  	Data server & 	A mobile FL scheme for 
			cooperative healthcare. &	The proposed model is simple with a lack of detailed analysis.  
			\\ \cline{2-9}&
			\cite{104} &	Healthcare cooperation &	DFL & - &	Data workers&  	Data workers &	A DFL architecture for healthcare with blockchain.  &	Performance of blockchain 	has not 	been verified.  
			\\ \cline{2-9}&
			\cite{106} &	Healthcare cooperation &	DFL & CNN &	Data centers &	Data centers &	A P2P network for 	for healthcare with FL. &	FL convergence has not been investigated.  
			\\ \cline{2-9}&
			\cite{108} &	Healthcare cooperation &	DFL & CNN &	Hospitals  &	Hospitals&  	A DFL framework for COVID-19 image classification. &	The simulations of blockchain is lacked.  
			\\ \cline{2-9}
			\hline
			
			\parbox[t]{1.5cm}{\multirow{18}{*}{\rotatebox[origin=c]{90}{FL for Smart Transportation }}} & 
			\cite{9} &	Vehicular traffic planning&  	HFL & - &	Vehicles &	Cloud server &	An FL approach for federated 
			traffic prediction. &	No detailed simulations are given.  
			\\ \cline{2-9}&
			\cite{112} &	Vehicular traffic planning &	HFL & RL&	Data owners &	Data center &	An FL-based scheme for traffic flow estimation. &	Numerical 
			simulations for privacy is lacked.  
			\\ \cline{2-9}& 
			\cite{114} &  	Vehicular traffic planning &	VFL & - &	Car agents &	FL server& 	An FL-based scheme for traffic collision avoidance tasks. &	Latency for FL simulations has not been investigated.  
			\\ \cline{2-9}&
			\cite{115} &	Vehicular traffic planning &	VFL & - &	UAVs  &	Data server &	A contract model for FLbased traffic control. &	The feasibility of the proposed model has not been evaluated.  
			\\ \cline{2-9}&
			\cite{116} &	Vehicular traffic planning &	DFL & NN &	Vehicles &	Vehicles &	A decentralized 
			FL approach with blockchain for vehicular networks.  &	The impacts of blockchain running on FL have not been investigated.  
			\\ \cline{2-9}&
			\cite{118} &	Vehicular resource management &	CFL & - &	Vehicles &	RSU &	An FL approach for resource allocation and power control in ITS. &	Learning latency has not been tested.  
			\\ \cline{2-9}&
			\cite{119} &	Vehicular resource management &	CFL&  DRL &	Vehicles &	BS &	A 	federated 
			resource 
			allocation 
			scheme 	for 
			V2X. &	Learning accuracy has not been investigated.  
			\\ \cline{2-9}&
			\cite{45} &	Vehicular resource management &	DFL & NN &	Vehicles &	RSUs &	A 	federated 
			resource management for MEC-based V2V. &	Privacy simulation of FL has 	not 	been verified.

			\\ \cline{2-9}
			\hline

	\end{tabular}}
\end{table*}

\begin{table*}
	\centering
	\caption{Taxonomy of FL-IoT applications (continued). }
	{\color{black}
		\label{Table:FL_Applications_next}
		\begin{tabular}{|P{0.5cm}|P{0.5cm}|P{1.7cm}|P{0.5cm}|P{0.7cm}|P{1.2cm}|P{1.5cm}|P{3.8cm}|P{3.8cm}|}
			\hline
			\textbf{Issue}& 	
			\textbf{Ref.} &	
			\textbf{Use case}&	
			\textbf{FL type}& 
			\textcolor{black}{\textbf{ML type}}& 
			\textbf{FL clients}& 	
			\textbf{FL aggregator}&
			\textbf{Key contributions}&
			\textbf{Limitations}
			\\
			\hline
			\parbox[t]{1.5cm}{\multirow{15}{*}{\rotatebox[origin=c]{90}{FL for UAVs}}} & 
			\cite{127} &		UAV 	path 
			control &	HFL & NN &	UAVs &	Central unit &	An FL-based scheme for federated UAV path control.  &	Training latency has 	not 	been analyzed. 
			\\ \cline{2-9}& 
			\cite{128} &	UAV 
			communications& 	HFL & - &	UAVs &	UAV &	An FL-based framework for 
			UAV power control and scheduling.  &	Privacy of UAV communications has 	not 	been taken 	into account. 
			\\ \cline{2-9}&
			\cite{129} &	UAV 
			beamforming design &	DFL & Extreme ML &	UAVs &	UAVs &		A 	federated 
			beamforming design for UAV communications. & 	UAV power control is not considered in FL design. 
			\\ \cline{2-9}&
			\cite{131}& 	UAV 
			communications &	VFL & CNN &	UAVs &	Data server &	An FL-enabled model for 
			illumination distribution 	in UAV communications. &  	The scalability of federated UAV network has not been explored.  
			\\ \cline{2-9}&
			\cite{133} &		UAV 	energy 
			management &	HFL & - &	UAVs &	Central server &	A federated UAV scheme for aerial air 	quality sensing.  &	UAV 
			communication latency has not been taken into account.   
			\\ \cline{2-9}&
			\cite{134} &	UAV 	security management &	HFL & NN &	UAVs  &	Data server& 	A 	federated scheme for attack detection in UAV networks. &	Convergence latency of FL algorithms 
			should 	be included.  
			
			\\ \cline{2-9}
			\hline
			
			\parbox[t]{1.5cm}{\multirow{15}{*}{\rotatebox[origin=c]{90}{FL for Smart city  }}} & 
			\cite{140} &		City 	data 
			management  &	VFL & DNN &	Vehicles &	BS &	A federated data learning 	for smart city.  &	The 	proposed smart city model is simple.
			\\ \cline{2-9}&  
			\cite{82} &		City 	data 
			sensing  &	DFL &NN &	IoT devices &	IoT devices& 	A federated sensing platform for smart cities. &	Convergence performance has not 	been verified.  
			\\ \cline{2-9}&
			\cite{36}& 		City 	video 
			analytics &	VFL & - &	Edge devices &	Cloud  &	A 	federated model for video analytics in smart cities. &	Issues related to privacy 	of videos have not been investigated.  
			\\ \cline{2-9}&
			\cite{148}& 		Smart 	energy 
			prediction &	HFL & NN &	Data centers &	Cloud &	A federated energy prediction in smart cities. &	Comparison of ML techniques in FL simulation has not been  performed.  
			\\ \cline{2-9}&
			\cite{149}& 		Energy 	load 
			prediction &	HFL & LSTM, DNN &	Smart meters  &	MEC 
			server &	A federated model for energy load prediction. &	Simulations for convergence 
			and learning latency have not been conducted.  
			\\ \cline{2-9}&
			\cite{151}& 	Power management& 	DFL & NN &	Electric providers& 	IoT users &	An FL-based scheme for power IoT networks. &	Learning efficiency, e.g., latency, has not been verified.  
			
			\\ \cline{2-9}
			\hline
			
		
	\parbox[t]{1.5cm}{\multirow{20}{*}{\rotatebox[origin=c]{90}{FL for Smart industry }}} & 
		\cite{154} &	Robotic management &	VFL& CNN &	Robotics &	Cloud 
		server  &	An FL-based scheme for robotic management. &	Data loss caused by communication has not been considered.  
		\\ \cline{2-9}&
		\cite{155} &	Federated cloud robotics &	HFL & NN &	Robotics &	Cloud server &	An imitation learning scheme for federated cloud robotics.  &	Data privacy and security have not been investigated.  
		\\ \cline{2-9}&
		\cite{158} &	Federated robotic arms &	HFL & RL &	Robotic arms &	Cloud server &	Federated reinforcement 
		learning 	for robotic arms.  &	Simulation 	is simple 	no details 	of FL results 	are reported.  
		\\ \cline{2-9}&
		\cite{161}& 	Industrial intelligence &	VFL & - &	Mobile devices &	Cloud server &	A 	secure federated model learning 	for industrial intelligence.  &	Learning  
		performances of FL have not been detailed.  
		\\ \cline{2-9}&
		\cite{162} &	Federated WiFi network &	HFL & - &	Access points &	Central server &	A federated compound TCP scheme in industry 4.0 WiFi networks.  &	Scalability for FL-based WiFi network has not been investigated.  
		\\ \cline{2-9}&
		\cite{163}& 	Secure industrial IoT &	DFL &DNN &	IoT devices &	IoT devices& 	An 	integrated FL-blockchain 
		scheme for industrial IoT. &	Performance of smart contracts has not been detailed.  
		\\ \cline{2-9}&
		\cite{401} &	Communication-efficient FL &	HFL& NN &	Edge devices&	Cloud server&	A  communication-efficient FL approach called CE-FedAvg for optimizing  communication rounds. &	The impact of lazy nodes on local training has not been considered.  
		\\ \cline{2-9}&
		\cite{403}&	Communication-efficient FL &	HFL& - &	Edge devices&	Model aggregator &	A solution for reducing communication overhead for FL-based edge IoT networks. &	Resource optimization for local training has not been investigated. 
		\\ \cline{2-9}&
		\cite{404}&	Network resource allocation&	HFL& - &	Edge devices&	Model aggregator& 	A fair network resource allocation framework for FL-based edge networks. &	Allocation of resource for model aggregation should be included.  
		\\ \cline{2-9}&
		\cite{405}&	Network resource allocation&	HFL& - &	Mobile users&	Edge server &	A network resource allocation scheme for FL-edge computing. &	Numerical simulations for privacy is lacked. 
		\\ \cline{2-9}&
		\cite{406}&	Network resource allocation&	HFL& DRL &	Mobile users&	Model owner &	A DRL-based approach for network resource optimization in FL-edge computing. &	The proposed FL scheme is simple and not scalable to multi-edge networks.
		\\ \cline{2-9}
		\hline
	\end{tabular}}
\end{table*}

%% file: Lessons_Learned.tex
\section{{Lessons Learned}}
\label{Sec:Lessons_Learned}
In this section, we summarize the key lessons learned from this survey, which thus provide an overall picture on the current research of FL-IoT services and applications. 
\subsection{Lessons Learned from FL-IoT Services}
Reviewing the state-of-the-art in the field, we find that FL plays an increasingly important role in facilitating intelligent IoT services in a wide range of applied domains, as highlighted as follows.
\subsubsection{FL Serving as an Alternative to IoT Data Sharing}
\textcolor{black}{FL-IoT can achieve {privacy-enhanced and scalable} data sharing in IoT networks among decentralized multiple parties without the need for direct data offloading to cloud servers or third-parties. From \cite{32}, it can be learned that FL is an efficient approach for federated data sharing among multiple clients (e.g., factories) in IoT tasks such as tensor mining. By using FL, the server only collects the ciphertext data and federates them into a tensor, while raw data are kept at the local factories, which thus protects data privacy for the tensor mining. Particularly, FL can be combined with other privacy techniques such as differential privacy \cite{34} to improve the privacy of local updates, by integrating it into gradient descent training for enabling secure and robust FL sharing. Moreover, the security of FL-based data sharing can be improved by combining with the blockchain technology, as shown in \cite{35}. In this context, the information of trained parameters can be appended into immutable blocks on the blockchain during the client-server communications.}

\subsubsection{FL for the Optimization of IoT Data Offloading and Caching}
Due to the high data distribution of large-scale IoT networks, FL can provide distributed AI solutions to support intelligent data offloading and caching at the network edge \cite{40}. The use of FL enables distributed data offloading by leveraging computation capabilities of IoT devices to achieve an overall offloading target, such as offloading latency minimization \cite{41}. Each FL client (such as a mobile device) acts as a local optimizer to solve an optimization problem formulated to achieve optimal offloading. Also, the federation of mobile users helps eliminate the need for a central server of offloaded data processing, instead of performing local training by using their own dataset. Then, each user exchanges the local updates for computing a global model for privacy and computing efficiency. FL can also offer innovative data offloading in vehicular networks \cite{45} in which vehicles and RSUs can share a common learning model to build a cooperative learning architecture, aiming to reduce learning costs with respect to agent selection and data sharing in task offloading.

In addition to that, the use of FL also helps overcome the challenges faced by traditional learning approaches in terms of high privacy concerns since data users may not trust the third-party server and thus hesitate to offload their private data for IoT data caching. In fact, FL is very useful to build proactive data caching schemes in edge computing without the need for direct access to user data \cite{49}. Such that, mobile users can download the AI model from a cache entity, such as an edge server, to perform local training before sending back the computed model to the server for aggregation in an iterative manner, for selecting the most popular files for caching. FL is also integrated with blockchain to build secure learning schemes for content caching in edge computing \cite{53}. IoT devices participate in the local training using their own datasets to learn the features of users and files, and share the gradient updates to the edge server for popular file estimation, aiming to improve the overall cache hit rate.
\subsubsection{FL for IoT Attack Detection}
The heterogeneity of IoT applications and services has become a main target of malicious adversaries that can attack AI/ML models integrated in IoT networks. FL has emerged as a strong alternative to perform distributed learning for IoT networks with the ability to detect a wide range of attacks and support network defense solutions \cite{56}. {Enabled by the privacy-enhancing features of FL}, federated attack detection and defense solutions can be realized where each IoT device joins to run an AI model, such as DNN, in order to retrain the threat model to fight against adversaries \cite{57}. The cooperation of multiple devices accelerates the learning process and improves learning accuracy while mitigating the risks of attack on the model learning. Moreover, FL also facilitates the detection of compromised IoT devices in federated IoT networks. In fact, with the sophistication of attacks and threats, it is challenging to detect them with current centralized AI approaches that often recognize attacks by making deviations from {user behavior profiles,} which suffers from high false alarm rate and long detection delay. FL comes as a natural solution to perform attack detection algorithms for distributed IoT networks \cite{64}. Each IoT can submit a detection profile obtained by local training to a security gateway where local updates are aggregated to build a common detection model for the IoT network. The involvement of a large number of IoT networks with diverse features and massive datasets enhances the learning accuracy for better attack detection efficiency.  To improve the scalability of detection of attacks, such as intrusion, line-speed and scalable FL approaches have recently introduced \cite{69}. Each IoT device runs a neural network for packet classification at the line speed of nearby switches in a scalable manner, while preserving privacy of network traces.
\subsubsection{FL for IoT Localization}
\textcolor{black}{FL can provide interesting solutions for enabling efficient and {privacy-enhanced localization services}, aiming to solve problems such as the lack of robustness due to the dynamics of mobile environments and privacy leakage bottlenecks caused by the centralized data processing architecture. As indicated in \cite{73}, FL can provide privacy for indoor localization services by forming a decentralized indoor localization scheme using the computational capability of distributed mobile devices without compromising sensor data. We also find that homomorphic encryption techniques are very useful to further improve privacy in FL training, by allowing for local update encryption which helps prevent the central server from guessing decryption result before  aggregation. This in turn provides high security for data training while ensuring a high location estimation accuracy. Furthermore, according to \cite{74}, FL is able to minimize the bias of location estimation with privacy guarantees by cooperating multiple users in the model learning using local fingerprint data. The usefulness of FL thus opens new opportunities for emerging localization services, such as localization in mobile indoor networks with global positioning system, mobile target tracking and navigation.}

\subsubsection{	FL for IoT Mobile Crowdsensing}
Traditionally, the reliance of a central server such as a single cloud to handle all sensing data is not a scalable solution, making it hard to cope with the massive volume of data from ubiquitous IoT devices. FL would be a very promising tool to accelerate the learning and training for crowdsensing models. Each client selects a learning strategy for solving its local sub-problem to ensure desired accuracy with lowest participation costs, while the central server builds a utility function by averaging local updates to offer rewards to the clients \cite{80}. Particularly, to overcome the challenges faced by conventional FL architectures with a central aggregator, which can incur high communication overhead and less robustness, decentralized FL has emerged as a promising solution for large-scale crowdsensing tasks \cite{82}. In this way, each IoT node only needs to compute the model based on its mini-batch instead of the whole datasets, which will reduce computation latency accordingly. Among decentralized technologies, blockchain is a strong candidate that can be combined with FL to decentralize the learning process in UAV-based crowdsensing services \cite{84}. FL allows UAVs to train local models using sensing datasets and share updates via the blockchain ledger for server communication and model combination in a secure and transparent manner. 
\textcolor{black}{\subsubsection{FL-based Techniques for Privacy and Security in IoT Services and Networks}
FL also appears as an attractive approach for enabling intelligent privacy and security services in IoT networks. As indicated by \cite{408}, FL is useful to enhance data privacy in IoT environments (e.g, vehicular networks). This is enabled by the fact that participants (e.g., vehicles) are allowed to train models locally with their own data without the need for sharing sensitive user data, which contributes significantly to protecting data privacy. To replace the centralized entity that is commonly used in FL systems, blockchain is an ideal candidate to build a decentralized FL-IoT system to solve issues of single-point failures by using  consensus protocols (e.g., shared mining) to synchronize and coordinate the FL training \cite{410}, \cite{413}. These solutions give insights into developing serverless learning architectures for security improvements in FL-IoT systems. }

\subsection{Lessons Learned from FL-IoT Applications}
In this sub-section, we discuss the key lessons acquired from the use of FL in various IoT applications.
\subsubsection{FL for Smart Healthcare}
\textcolor{black}{FL can provide a number of efficient solutions for smart healthcare and potentially reshapes the current intelligent healthcare systems by proving AI functions for supporting healthcare services while improving user privacy and reducing low latency with the cooperation of multiple entities such as health users and healthcare providers across medical institutions \cite{87}. It can be learned from \cite{88} that FL can enable flexible and privacy-preserving EHRs management in healthcare operations, by building intelligent EHRs systems with the cooperation of {multiple medical institutions} and a powerful server such as the cloud. We also find that a few fully decentralized FL approaches \cite{91} can provide decentralized optimization and stochastic gradient tracking by using the cooperation of hospitals with a decentralized stochastic gradient algorithm to accelerate the convergence rate. Moreover, based on \cite{99}, personalized FL is necessary for mitigating data heterogeneities in distributed health IoT networks while attaining high-quality personalized models. FL also allows each IoT device to choose offloading of its computationally intensive tasks to edge gateways which execute the learning models before sending the updates to the cloud for combination. This realizes a cooperative network of cloud, edge, and IoT devices for healthcare applications with the help of differential privacy and homomorphic encryption techniques for security improvement.}

\subsubsection{FL for Smart Transportation}
Recently, FL has been introduced to bring AI functions to the network edge to empower smart transportation, involving a number of participants, such as vehicles, to collaboratively train globally shared AI models without the need for long data transmission and compromising user privacy. Several possible applications of FL in smart transportation can be vehicular traffic planning and vehicular resource management. In fact, FL can be used to replace traditional centralized ML approaches in traffic prediction tasks \cite{112,115} by running ML models directly at the edge devices, e.g., vehicles, based on their datasets such as road geometry, traffic flow and weather. The use of massive data from multiple vehicles and huge computation capability of all participant helps provide better traffic prediction outcomes, which cannot be met by using centralized ML techniques with less dataset and limited computation. Moreover, by combining with blockchain, FL is useful to build decentralized traffic planning solutions for vehicular systems \cite{116}. In this regard, each vehicle acts as an FL client to run an ML model and exchange computed updates together via a blockchain ledger while verifying their corresponding rewards. Further, FL has the potential to facilitate resource management strategies for vehicle-to-vehicle (V2V) networks \cite{118}. FL along with advanced AI techniques such as DRL \cite{120} is promising to build federated intelligent resource allocation strategies, in order to maximize the sum capacity of vehicular users with respect to latency and reliability conditions. Vehicles as DRL agents can collaborate to run distributed DNN algorithms for optimal mode selection and resource allocation, while the BS aggregates the updates offloaded from users to build undirected graphs using channel gain information.

\subsubsection{FL for UAVs}
In UAV networks, due to the high mobility and high altitude of UAVs, it is challenging to ensure the continuous communications between UAVs and BSs with respect to dynamic aerial environmental conditions to perform intelligent UAV tasks. The use of centralized AI/ML approaches in such scenarios may not be an ideal solution, especially when having to transmit a large amount of data over the aerial links. FL can provide better learning solutions for intelligent UAVs networks by using a cooperation of multiple UAVs without transferring raw data to BSs and compromising data privacy \cite{127}. The use of FL would mitigate the data volume transferred over the aerial environment and thus reduce communication delays and privacy concerns, while increasing training speed of the global model due to the employment of computing resources of all UAVs. Moreover, FL enables UAVs to collaboratively train AI models only by using their partial illumination data \cite{131}, which potentially reduces transmission power and improves privacy. In this way, UAVs have more flexible solutions to adjust its serving position and user association for communication power savings. FL also makes federated UAV management much more flexible and with privacy awareness \cite{133}. Each UAV acts as an FL client to monitor the air quality index in its coverage area without revealing raw data by cooperating UAVs with a central server which is responsible for running a light-weight DenseMobileNet model based on haze features offloaded from distributed UAVs. For security protection in UAV networks, FL is combined with reinforcement learning to implement defense strategies against jamming attack \cite{137}, by cooperating learning updates from FL models at UAVs with a Q-learning table to determine optimal UAV paths for minimal security risks. 

\subsubsection{FL for Smart City}
\textcolor{black}{To realize smart cities, AI/ML techniques have been widely adopted to provide intelligence properties thanks to their ability to handle real-time big data generated from sensors, devices, and human activities \cite{139}. However, most proposed AI-based smart city solutions rely on a centralized learning architecture on a data center, such as a cloud server, which is obviously not scalable to the rapid expansion of smart devices in smart cities. FL offers more attractive features for enabling decentralized smart city applications with high privacy levels and low communication delays \cite{140}. FL is also important for structuring data streams from ubiquitous IoT devices that act as FL clients for performing local learning without sharing their data with external third-parties \cite{142}. This would reshape the current forms of smart cities by providing new services such as smart urban communication, social economy sharing, social activity monitoring, and interconnection of global citizens. Moreover, we also realize that FL can enable intelligent solutions for smart grid, a critical component of smart city ecosystems, by offering management and energy transmission solutions in a decentralized manner {with privacy improvement} \cite{148}.}
\subsubsection{FL for Smart Industry}
FL can also provide viable solutions to realize intelligence for industrial smart systems with many applied domains such as robotics and Industry 4.0 without data exchange and privacy leakage. Following \cite{154}, FL is an attractive approach to coordinate the data learning among robotics for industrial tasks, e.g., traffic routing, without unpredictable network transmission delays. Edge computing is also integrated with FL to realize a collaborative learning among robotic arm devices \cite{158}. Due to the complexity and dynamics, each device runs a separate ML model, e.g, reinforcement learning, for determining its own control policy, and then shares mature policy model parameters to a cloud server for aggregation. FL also plays an important role in supporting distributed intelligent industrial applications in Industry 4.0 \cite{161}. Particularly, decentralized FL in Industry 4.0 can be realized by combining with the blockchain technology which can provide high degrees of security with immutable ledgers and smart contracts that can provide authentication for learning interactions in FL implementation \cite{163}. Furthermore, to enable intelligent and privacy-enhanced edge-based  industrial IoT applications, many efficient FL solutions have been introduced to solve communication  issues by using update reconstruction techniques \cite{401}, \cite{402} or relevance detection during the model update \cite{403}.  Network resource management is another critical issue in industrial FL-IoT system design that has been solved via resource allocation \cite{404} and stochastic training optimization \cite{406}. 

%% file: Challenges_Future-Directions.tex
\section{Research Challenges and Directions}
\label{Sec:Challenges_Future-Directions}
As discussed in the above sections, FL demonstrates its increasingly significant role in empowering IoT services and applications. Despite its great potential, the extensive survey also reveals several critical research challenges to be considered for future FL-IoT system implementation. We here analyze several key challenges, concerning FL such as threats, performance issues, resource management, and heterogeneity issues in IoT networks. Several possible research directions for these challenges are also discussed. 
\subsection{\textcolor{black}{Security and Privacy Issues in FL}}
Although FL can provide privacy protection for distributed IoT systems where the sharing of raw IoT data is not required in the learning process, FL still remains several security and privacy vulnerabilities from both learning clients and server sides \cite{165}. For example, at the client side, an adversary can modify data features or inject an incorrect subset of data in the original dataset to embed backdoors into the model, aiming to adjust the training objective of local clients. This is also called as backdoor poisoning attacks which also poison local model updates at the clients before offloading to the server \cite{166}. Meanwhile, the central server may contaminate the aggregated local updates and deploy attackers to steal the training data from gradients in a few iterations \cite{168} because the fact that the gradients of the weights are the inner products of the error of learning layers and features. Consequently, the global updating can reveal extra information about personal features of local training data which thus poses user privacy at risks. This issue becomes more critical in the IoT networks where user information such as user preference, home addresses must be preserved in applied intelligent domains with FL. Without a protection mechanism, FL becomes a security and privacy bottleneck in intelligent IoT systems and thus makes it challenging to attract users in the collaborative data training. 

Under these contexts, developing innovative solutions to cope with threat challenges is of paramount importance for safe FL systems. Perturbation techniques \cite{168} such as differential privacy or dummy can be used to protect training datasets against data breach, by constructing composition theorems with complex mathematical solutions. \textcolor{black}{As an example, differential privacy is applied in \cite{169} by inserting artificial noise (e.g., Gaussian noise) to the gradients of neural network layers to preserve training data and hidden personal information against external threats while the convergence property is guaranteed. This solution would ensure that the server or malicious users cannot learn much additional information of user samples from the received messages under any auxiliary information and attack. Similarly, an anonymous and privacy-preserving FL scheme is introduced in \cite{500} for the mining of industrial big data. To mitigate privacy leakage, fewer parameters are shared between the server and each participant while having less impact on the model accuracy. Differential privacy is then applied on shared parameters with a Gaussian mechanism to provide strict privacy preservation on a proxy server as the middle layer between the server and all the participants. Moreover, the study in \cite{170} proposes a secure aggregation scheme for safe FL systems, in order to provide the strongest possible security with respect server-mediated, unauthenticated network conditions. The core idea is to use a double-masking structure that can protect clients by encrypting local updates and key sharing among users and the server for achieving fair data verification and key agreement during the aggregation. \textcolor{black}{The further privacy protection in FL-IoT is still an ongoing research topic, and new innovative solutions and techniques are desired to improve privacy for FL-based IoT systems from both client (e.g., IoT device) and communication (e.g., over the wireless server-client links) perspectives.}  } 
\subsection{\textcolor{black}{Communication and Learning Convergence Issues of FL-IoT}}
\textcolor{black}{Another challenge raised from FL-IoT implementation is its limited performance in terms of communication and learning convergence. In fact, communications in FL training in both uplinks and downlinks are highly sensitive {due to the unbalanced and non-IID data} since each training data at each client is different in size and distribution owing to different sensing environments. Further, when the number of clients grows exponentially, direct communications between numerous clients and a server for offloading updates become infeasible due to the increasing workload on the network channels \cite{171}. In addition  to that, the convergence of FL algorithms in IoT systems is not always ensured \textcolor{black}{due to the heterogeneous training capabilities of different IoT devices} and training data complexity. }  Indeed, the well-known FL algorithm FedAvg makes limited assumption that all IoT device clients join in each communication round for FL training \cite{172}, which is often not feasible in realistic IoT scenarios due to device connection loss or running out of battery.  Moreover, the convergence speed of current FL algorithms is constrained due to the use of first-order gradient descent in loss function minimization \cite{175}. 

Several possible solutions have been proposed to solve issues related to FL performances. The research in \cite{173} proposes a new efficient communication protocol that is able to compress uplink and downlink communications while remaining high robustness to {the increased number of clients}, and data distribution. These properties can be achieved by using a combination of sparsification, ternarization, error accumulation, and optimal Golomb encoding techniques for uplink compression and speeding up parallel training in the global server without compromising learning convergence. \textcolor{black}{Moreover, a new optimization algorithm is proposed in \cite{501} for FL-based IoT networks, called FetchSGD, that can train high-quality models for communication efficiency. At each communication round, clients compute a gradient based on their local data, then compress the gradient using a data structure called a Count Sketch before sending it to the central aggregator. The aggregator maintains momentum and error accumulation count sketches, and the weight update applied at each round is extracted from the error accumulation sketch. In this way, the proposed scheme reduces the amount of communication required each round, while still conforming to the training quality requirements of the federated setting.} To improve the convergence rate of FL algorithms, a new FL design called Momentum federated learning is presented in \cite{174}. A momentum gradient descent (MGD) approach is integrated at the central server to minimize the loss function with less computation due to optimized learning parameters, compared to traditional FL algorithms with first-order gradient descent.
\subsection{Resource Management in FL-IoT}
The concept of FL-IoT mostly relies on scalable data parallelism and on-device training at IoT devices before the aggregation of learning parameters at a global server. To achieve a synchronous update at the server, all IoT devices need to devote their storage and computing resources for their data training. Unfortunately, this requirement is not always met due to the resource constraints of certain IoT devices with weak computation capacities \cite{176}, which thus can cause significant delays to the synchronized parameter aggregation at the server. Moreover, training deep learning models such as DNN directly on IoT devices may not be achieved due to the requirement of much CPU frequency and battery for solving training tasks, especially training with image and audio data \cite{177}. Compared to edge devices, resources of IoT devices are still very limited for large AI training.  Therefore, optimizing one-device AI/ML models would be a viable solution that can mitigate the computational burden posed on IoT devices.

To facilitate resource management in on-device FL training, several approaches have been provided. The work in \cite{178} proposes a resource-aware FL architecture on mobile devices which are used to train neural networks by taking information of computation resource into consideration. To be clear, a soft training technique is introduce for accelerating the training of stragglers with weak computational capabilities. This can be done by allowing them to train partially the model through masking a particular number of resource-intensive neurons during their local training stage, but being recovered in the parameter aggregation stage without compromising the overall model convergence. For optimized on-device AI models, an improved DNN architecture called DeepRebirth is designed in \cite{179} for mobile AI training.  Particularly, a streamlined slimming model is integrated with the consecutive non-tensor and tensor layer which has the potential to speed up the training and model execution and thus can mitigate running time and save memory resources. Simulations also confirm a high training performance with 3x training speed and 2.5x memory saving while only 0.4\% performance drop on top-5 accuracy is observed via the ImageNet validation dataset.
\textcolor{black}{Another possible solution for resource management in FL is \cite{502} where data-importance and compute communication-aware resource management algorithms are proposed to optimize training accuracy, fairness and convergence time. The attention is focused on the resource management problem of sampling clients in each round taking into account communication, compute and statistical heterogeneity with the goal of reducing convergence time. To be clear, importance sampling and rank ordering based algorithms were developed that allow prioritization of clients based on their resources in the presence of non-I.I.D data across clients. Performance evaluation on a supervised image classification task on benchmark datasets showed significant reduction in the overall training time without loss of performance in the model training.} 

\subsection{Feasibility of Deploying AI Learning Functions on IoT Sensors}
Another possible challenge is the feasibility of running AI functions on IoT sensors. Due to the constraints of hardware, memory, and power resources, certain IoT sensors cannot join to train a full-size AI model \cite{challenge1}. In fact, advanced ML algorithms often takes a large amount of memory and power for model training and the storage of model parameters and training variables. For example, training relatively simple image classification models such as ResNet-50 still requires a number of CPU cycles and megabytes of memory space \cite{challenge2}. Moreover, the high communication cost caused by AI training is also a crucial bottleneck of on-device FL implementation. Indeed, the model exchange between IoT sensors and the server incurs communication overhead which scales up with the model size \cite{challenge3}. This would discourage IoT sensors from participating in training large models for running on-device FL algorithms. \textcolor{black}{In addition to that, how to address the energy consumption issue in FL-IoT systems, especially in the light-weight mobile devices with less battery sources is another critical challenge. In this context, it is important to achieve effective and efficient local training on local devices for energy savings while the training quality should be ensured.}

Several possible solutions should be considered to facilitate on-device AI learning for FL-IoT systems. One direction is to improve AI hardware usage on IoT sensors. The work in \cite{challenge4} introduces a software-based deep learning accelerator to support AI/DL training on mobile hardware. The key idea is to use a set of heterogeneous processors (e.g., GPUs) where each computing unit exploits distinct computational resources for processing different inference phases of DL models. This aims to optimize hardware usage for data training without {compromising performance accuracy}, enabled by the control of two algorithms, i.e., runtime layer compression and deep architecture decomposition. Simulation results demonstrates its superior performance in terms of low execution time and energy consumption in AI hardware running, compared to cloud offloading-based approaches. This research is promising and {likely to enable developing} sensor processing and mobile AI inference, which would enable on-device FL implementation at scale. \textcolor{black}{Additionally, a scheme called Tiny-Transfer-Learning (TinyTL) is proposed in \cite{503} for memory-efficient on-device sensor learning. TinyTL freezes the weights while learning only the bias modules, thus there is no need to store the intermediate activations which reduces  the memory footprint. To compensate for the capacity loss, a memory-efficient bias module is integrated which improves the model capacity by refining the intermediate feature maps of the feature extractor with a small memory overhead. Through simulations on image classification datasets, this approach can achieve the same level accuracy compared to fine-tuning the full Inception-V3 while reducing the training memory footprint by a factor of up to 12.9, which has the potential to allow for on-device AI implementation, including FL settings on IoT sensor devices. } In terms of communication cost in on-device FL training, the authors in \cite{challenge5} propose a light-weight model training strategy, by using a concept of model output exchange instead of model parameter exchange. Such that, only model outputs are exchanged between IoT devices and the server, which potentially solves communication latency issues caused by the increase of model size. This new method can achieve a significant latency reduction up to 99\% compared to existing FL algorithms, for the similar training accuracy performance under non-IID data distributions. \textcolor{black}{Moreover, to solve the energy consumption issue in FL-IoT systems, a number of advanced techniques are integrated, including gradient sparsification, gradient quantization, weight quantization, and dynamic batch sizes in the FL training procedure to mitigate energy costs for 5G mobile devices \cite{602}. A trade-off analysis is also provided which gives insights into how to balance the energy consumption for local computing for model training and wireless communications for model exchange, aiming to boost the overall energy efficiency.}

\textcolor{black}{\subsection{Standard Specifications}
The introduction of vertical FL-IoT use cases in future intelligent networks imposes major architectural changes to current mobile networks in order to simultaneously support a diverse variety of stringent requirements (e.g. autonomous driving, e-healthcare, etc.) In such a context, network standards and elements play an important role in deploying FL-IoT ecosystems at large-scale due to the reliance of other important computing services such as edge/cloud analytics server and edge-IoT communication protocols. Such elements have the power to fill the gaps in the FL-IoT ecosystem to make it sharper and technologically visible. For example, the Industry Specification Group (ISG) of the European Telecommunications Standards Institute (ETSI) has released an initiative called ETSI Multi-access Edge Computing (MEC) \cite{414}, which aims to leverage seamless and open edge computing and communication frameworks for integrating various edge computing-based applications originating from the vendors, developers and third-party service providers. All authentic operators can run their specific network domains at the edge of deployed MEC. This initiative could be a significant element to the FL-IoT systems where FL aggregation can be done by the support of computing services offered by MEC. IoT data training can also be stored and processed off-line at the edge nodes which will also facilitate services like video analytics, augmented reality, data caching, and content delivery. Moreover, to enable FL-IoT services and applications, standards for edge-IoT communication protocols are essential.  For example, OPC-UA has been applied widely  for edge-IoT scenarios \cite{415}. This Open Platform Communication standard is designed to support the platform independent Unified Architecture for edge-IoT services. Another protocol is MODBUS which is the communication standard for connecting industrial electronic equipment. MODBUS comes with a variety of protocols such as remote terminal unit (RTU), TCP/IP, UDP, Plus, Pemex and Enron. It relies on a mesh networking architecture and is able to communicate with the supervisory control and data acquisition systems over the industrial, scientific and medical radio bands. Among wireless protocols, Wi-Fi is one of the most prevalent IoT communication protocols, allowing for connections between IoT devices and computing servers such as edge nodes in FL-IoT systems. Very recently, the IEEE 802.11 working group has initiated discussions on releasing the next generation of Wi-Fi standard, referred to as IEEE 802.11be Extremely High Throughput \cite{416}, which can meet the peak throughput requirements set by upcoming IoT applications in the 5G/6G era. These standards are expected to support service providers in deploying intelligent IoT services at the network edge with FL-IoT components and devices.} In summary, the research challenges and directions are highlighted in Table~\ref{Table:Challenge}.

\begin{table*}
	\centering
	\caption{\textcolor{black}{Summary of research challenges and possible directions for FL-IoT.	} }
	\label{Table:Challenge}
	{\color{black}
		\begin{tabular}{|P{2.6cm}|P{7cm}|P{7cm}|}
			\hline
			{\cellcolor{blue!25}\centering \textbf{Challenges}}& 	
			{\cellcolor{blue!25}\centering \textbf{Description}}& 
			{\cellcolor{blue!25}\textbf{Possible directions}	}
			\\
			\hline
			\multirow{8}{*}{\makecell{Privacy and Security \\ Issues in FL}} &
			\begin{itemize}
				\vspace*{-1mm}
				\item 	At the client side, an adversary can modify data features or inject an incorrect subset of data into the original dataset to embed backdoors into the model \cite{166}. 
				\item	The central server can contaminate the aggregated local updates and deploy attackers to steal the training data from gradients in a few iterations \cite{168}.
			\end{itemize}& 
			\begin{itemize}
				\vspace*{-1mm}
				\item 	\textcolor{black}{	Differential privacy is applied in \cite{169}, \cite{500} by adding artificial noise (e.g., Gaussian noise)  to learning gradients to preserve training data and hidden personal information against external threats.}
				\item	A secure aggregation scheme is proposed in \cite{170} for safe FL systems, in order to provide the strongest possible security for data aggregation. 
				
			\end{itemize}
			\\ \cline{2-3}
			\hline
			\multirow{8}{*}{\makecell{Communication and \\ Learning  Convergence \\ Issues in FL-IoT}} & 
			\begin{itemize}
				\vspace*{-1mm}
				\item 	Communication in FL training is highly sensitive due to non-IID data distribution and increasing numbers of clients in distributed settings \cite{171}.
				\item	The convergence rate of traditional FL algorithms is limited due to the issues of heterogeneous data and resources of distributed IoT devices \cite{172}.
			\end{itemize}& 
			\begin{itemize}
				\vspace*{-1mm}
				\item 	A new protocol is proposed in \cite{173} for fast uplink and downlink communications with increasing numbers of clients and non-IID data. Also, an algorithm called FetchSGD is proposed in \cite{501} for FL-based IoT networks that can train high-quality models for communication efficiency.
				\item	A Momentum FL scheme is designed in \cite{174} to minimize the loss function and improve the convergence rate with less computation.  
			\end{itemize}
			\\ \cline{2-3}
			\hline
			
			\multirow{8}{*}{\makecell{Resource Management \\ in FL}} & 
			\begin{itemize}
				\vspace*{-1mm}
				\item 		Resource requirements for FL training are not always met due to weak computation capabilities at certain IoT devices \cite{176}. 
				\item 	Training deep learning models directly on IoT devices may not be achieved  due to limitations in CPU frequency and battery capacity \cite{177}.
			\end{itemize}& 
		\vspace*{-1mm}
			\begin{itemize}
				\item 		A resource-aware FL architecture is proposed in \cite{178} to train neural networks with computation resource awareness.
				\item 	An improved DNN architecture is designed in \cite{179} for optimized mobile AI model training.  Moreover, communication-aware resource management algorithms are proposed in \cite{502} to optimize training accuracy, fairness and convergence time.
			\end{itemize}
			\\ \cline{2-3}
			\hline
			
			\multirow{8}{*}{\makecell{Deploying AI Functions \\on IoT Sensors}} & 
			\begin{itemize}
				\vspace*{-1mm}
				\item 	Due to constraints on hardware, memory, and power resources, certain IoT sensors cannot join to train a full-size AI model \cite{challenge1}.
				\item 	The high communication cost and energy consumption caused by AI training places limits on on-device FL implementation \cite{challenge3}.	
			\end{itemize}& 
			\begin{itemize}
				\vspace*{-1mm}
				\item 	A software-based deep learning accelerator is introduced in \cite{challenge4} to optimize hardware usage for AI/DL training on mobile hardware. \textcolor{black}{A memory-efficient on-device sensor learning solution is also introduced \cite{503} which can achieve high on-device training accuracy with less memory overhead.}
				\item	\textcolor{black}{Energy-aware model training strategies are important for mitigating significant network latency \cite{challenge5} and optimizing energy costs in FL-based mobile devices \cite{602}.} 
			\end{itemize}
			\\ \cline{2-3}
			\hline
			
			\multirow{10}{*}{\makecell{Standard Specifications \\ in FL-IoT}} & 
			\textcolor{black}{\begin{itemize}
				\vspace*{-1mm}
				\item 	The introduction of vertical FL-IoT use cases in future intelligent networks imposes major architectural changes to current mobile networks to simultaneously support a diverse variety of stringent requirements.
				\item 	Network standards and elements play an important role in deploying FL-IoT ecosystems at large-scale due to the reliance of other important computing services such as edge/cloud analytics server and edge-IoT communication protocols.
			\end{itemize}}& 
			\textcolor{black}{\begin{itemize}
				\vspace*{-1mm}
				\item The initiative called ETSI Multi-access Edge Computing (MEC) has been released \cite{414} which leverages seamless and open edge computing and communication frameworks for integrating various MEC-based applications from the vendors and service providers. 
				\item The IEEE 802.11 working group has initiated discussions on releasing the next generation of Wi-Fi standards, referred to as IEEE 802.11be Extremely High Throughput \cite{416}, which can meet the peak throughput requirements set by future IoT applications. 
			\end{itemize}}
			\\ \cline{2-3}
			\hline
	\end{tabular}}
\end{table*}

%% file: Conclusion.tex
\section{Conclusions} 
\label{Sec:Conclusion}
FL is an emerging distributed AI approach that has sparked great  interest to realize privacy-enhancing and scalable IoT services and applications. In this article, we have explored the opportunities brought by FL to facilitate IoT networks through a state-of-the-art survey and extensive discussions based on the emerging studies in the field. This work is motivated by the lack of a comprehensive survey on the use of FL in IoT. To bridge this gap, we have first introduced the recent advances in FL and IoT and {provided insights into their integration}. We have then provided an updated survey on the application of FL in key IoT services, namely IoT data sharing, data offloading and caching, attack detection, localization, mobile crowdsensing, and IoT privacy and security. {Subsequently}, we have paid an attention to the discussion of latest developments of the integrated FL-IoT applications in various significant use-case domains, including smart healthcare, smart transportation, UAVs, smart city, and smart industry. From the extensive survey, several main lessons learned have been also summarized and analyzed. Finally, we have identified a few key research challenges and possible directions for further investigation. We believe that this article will stimulate more attention in this emerging area, and encourages more research efforts toward the full realization of FL-IoT.